\documentclass[prc,aps,nofootinbib,showkeys,showpacs,twocolumn]{revtex4}
\usepackage{amssymb}
\usepackage{amsmath}
\usepackage{graphics}
\usepackage{color}
\usepackage[usenames,dvipsnames]{xcolor}
\usepackage{graphicx}

\usepackage{rotating}

\begin{document}
% --------------------------------------------------
\title{Static response, collective frequencies and ground state thermodynamical properties of
spin saturated two-component cold atoms and neutron matter}
\date{\today}
\author{A. Boulet}\email{boulet@ipno.in2p3.fr}
\affiliation{Institut de Physique Nucl\'eaire, IN2P3-CNRS, Universit\'e
Paris-Sud, Universit\'e Paris-Saclay, F-91406 Orsay Cedex, France}
\author{D. Lacroix}\email{lacroix@ipno.in2p3.fr}
\affiliation{Institut de Physique Nucl\'eaire, IN2P3-CNRS, Universit\'e
Paris-Sud, Universit\'e Paris-Saclay, F-91406 Orsay Cedex, France}

\begin{abstract}
The thermodynamical ground-state properties and static response in both cold atoms at or close to unitarity and neutron matter
are determined using a recently proposed Density Functional Theory (DFT) based on the $s$-wave scattering  length $a_s$, effective range $r_e$, and unitary gas limit. In cold atoms, when the effective range may be neglected, we show that the pressure, chemical potential, compressibility modulus and sound velocity obtained with the  DFT are compatible with experimental observations or exact theoretical estimates. The static response in homogeneous infinite systems is also obtained and a possible influence
of the effective range on the response is analyzed.
The neutron matter differs from unitary gas due to the non infinite scattering length and
to a significant influence of effective range  which affects all thermodynamical quantities as well as the static response.
In particular, we show for neutron matter that the latter response recently obtained in Auxiliary-Field Diffusion Monte-Carlo (AFDMC) can be qualitatively reproduced when the $p$-wave contribution is added to the functional.
Our study indicates that the close  similarity between the exact AFDMC  static response
and the free gas response might stems from the compensation of the $a_s$ effect by the effective range and $p$-wave contributions.
We finally consider the dynamical response of both atoms or neutron droplets  in anisotropic traps.
Assuming the hydrodynamical regime and a polytropic equation of state, a reasonable description of the radial and axial
collective frequencies in cold atoms is obtained.
Following a similar strategy, we estimate the equivalent collective frequencies of neutron drops in anisotropic traps.
\end{abstract}



\maketitle

\section{Introduction}
The static and/or dynamical responses of many-body interacting systems give important
information on the interaction between their constituents. In the last decade, important progresses have been
made on the understanding of diluted cold atoms properties \cite{Blo08,Gio08,Chi10,Zwe11} with varying $s$-wave scattering length,
eventually reaching the unitary gas (UG) limit for which $|a_s k_F|^{-1} \rightarrow 0$.  Due to the very large scattering length in neutron matter,
these progresses directly impact nuclear physics and offers the possibility to address the nuclear  many-body problem from a new
perspective \cite{Lac16,Tew16,Kon17}. Another interesting progress in nuclear physics is the possibility to perform exact calculations
based on Quantum Monte-Carlo or other many-body techniques \cite{Car12,Gan15}.  However, contrary to the cold-atom case,
while many efforts have been made to study the properties of nuclear Fermi systems in their ground states, very little is known from exact theories
away from it. Very recently, the exact static response of neutron matter at various densities has been
studied in Refs. \cite{Bur16,Bur17}.
These benchmark calculations give new pieces of information on neutron matter and stringent constraints for other many-body approaches like the
nuclear  density functional theory (hereafter called Energy Density Functional [EDF]). In particular, it was noted in Ref. \cite{Bur17}, that the
empirical Skyrme EDF leads to static response having significant differences with the exact
case.

The recent work on ab-initio static response in neutron system together with the recently developed
density functional proposed in Refs. \cite{Lac16,Lac17}
that makes a clear connection between cold atoms and neutrons systems are the original motivations of this work. Starting from this functional,
we first analyze the ground state thermodynamical properties of both cold atoms and neutron matter, some of them being directly linked to the
static response. In particular, we underline the key role played in neutron matter by the effective
range.
We finally conclude the work by an exploratory study of the collective response of cold atoms and neutron droplets in an anisotropic trap.

\section{Introduction of the functional}

The functional proposed in Refs. \cite{Lac16,Lac17} may be written as:
\begin{align}
  \frac{E}{E_\mathrm{FG}} &\equiv \xi(a_sk_F,r_ek_F) , \nonumber \\
  &= 1 - \frac{U_0}{1-(a_sk_F)^{-1}U_1} \nonumber \\
   &+ \frac{R_0(r_ek_F)}{\left[1-R_1(a_sk_F)^{-1}\right]
  \left[1-R_1(a_sk_F)^{-1}+R_2(r_ek_F)\right]},
\label{eq:func1}
\end{align}
where $ \xi(a_sk_F,r_ek_F) $ can be understood as a generalization of the Bertsch parameter for finite
$s$-wave scattering length $a_s$ and effective range $r_e$.  $E_\mathrm{FG}$ is the free Fermi--gas (FG) energy given by:
\begin{eqnarray}\label{eq:EFG-def}
\dfrac{E_\mathrm{FG}}{N} = \dfrac{3}{5} \dfrac{\hbar^2k_F^2}{2m} \equiv \mathcal{E}_\mathrm{FG},
\end{eqnarray}
where $N$ is the number of particles in a unit volume $V$.
In the present article, we will consider a infinite  spin equilibrated system  formed of one particle type only (cold atoms or neutrons).
Then, the  Fermi momentum $k_F$ is linked to the density $\rho $ through  $\rho  = k_F^3/(3\pi^2)$.
The different parameters are fixed by imposing specific asymptotic limits either in the low density
regime $(a_s k_F) \rightarrow 0$
and/or unitary limit $-(a_s k_F) \rightarrow +\infty$. Values of parameters obtained in Ref.
\cite{Lac17} are recalled in table \ref{tab:funcparam}.

 \begin{table}
%\begin{tabular}{|c|c| c|}
\begin{tabular}{|c|c|}
\hline
& Low density+Unitary gas  \\ %&   Low density only \\
\hline
\hline
%$U_0$  & $1-\xi_0 = 0.62400$ & $\frac{20}{27\pi^2}C^{-1}_0 = 0.674192$ \\
%$U_1$ & $ \frac{9\pi}{10}(1-\xi_0)=1.76432 $  &   $ \frac{2}{3 \pi} C^{-1}_0=1.90623$ \\
%$ R_0$ & $\eta_e = 0.12700$ & $ U_0R_2 = 0.192775$ \\
%$R_1$ &  $\sqrt{6\pi\eta_e} = 1.54722$  & $U_1 = 1.90623$\\
%$R_2$ & $-\delta_e / \eta_e = 0.43307$ & $\frac{1}{10 \pi}  C^{-1}_0=0.285935 $ \\
$U_0$  & $1-\xi_0 = 0.62400$  \\
$U_1$ & $ \frac{9\pi}{10}(1-\xi_0)=1.76432 $   \\
$ R_0$ & $\eta_e = 0.12700$  \\
$R_1$ &  $\sqrt{6\pi\eta_e} = 1.54722$  \\
$R_2$ & $-\delta_e / \eta_e = 0.43307$ \\
\hline
\end{tabular}\caption {Values of the different parameters entering in the functional.
The parameters $\xi_0$ (Bertsch parameter), $\eta_e$ and $\delta_e$ are defined at the unitary limit  $(a_s k_F)^{-1}=0$ from the constraint
$  \xi(a_sk_F\to\infty,r_ek_F) \equiv \xi_0 + (r_ek_F)\eta_e + (r_ek_F)^2\delta_e+ \cdots$
with the values obtained in Ref.  \cite{Ku12,For12}: $\xi_0=0.376$, $\eta_e=0.127$ and $\delta_e=-0.055$.
%For the sake of completeness, we also give in the left hand side the set of parameters obtained when the functional
%is adjusted to reproduce only the low density limit provided by the Lee-Yang formula \cite{Lee57,Bis73}:
%$E/E_{\rm FG} = \dfrac{10}{9 \pi} (a_sk_F) + \dfrac{1}{6 \pi} (a_sk_F) (r_e k_F) + \dfrac{5}{3}C_0 (a_s k_F)^2+ \cdots$ with
%$C_0 = \dfrac{4}{35\pi^2}(11-2\ln 2) $. Note that in the low energy case, the original version of the functional was used \cite{Lac16}
%where only 3 parameters out of the 5 are independent.
}
\label{tab:funcparam}
\end{table}

\section{Application to cold atoms}

The physics of cold atoms has attracted lots of attention in the past decades \cite{Blo08,Gio08, Chi10}.
One advantage in this case is the possibility to adjust the $s$-wave scattering length at will while, in some cases,
effects of effective range and other channels can be neglected.  Assuming first that $r_e =0$, we end-up with the simple functional that depends solely
on the Bertsch parameter $\xi_0$ in the unitary regime. The great simplification of the DFT compared to other many-body theories stems from the
fact that any quantity that could be written as a set of derivatives of the energy with respect of the density can be obtained in a straightforward manner.   An illustration has been given in Ref. \cite{Lac16} with the Tan contact parameter \cite{Tan08_1,Tan08_2,Tan08_3}. Below, we give other examples with
ground-state thermodynamical quantities.

\begin{table*}
  \centering
%  \begin{turn}{90}
%  \begin{tabular}{|c||c|c|c|c|c|}
  \begin{tabular}{|c||c|c|c|c|}
\hline
\hline \rule[-0.6cm]{0.cm}{1.5cm}
                          $X/X_\mathrm{FG}$
                       &  $E/E_\mathrm{FG}$
                       &  $\mu/\mu_\mathrm{FG} = \mu\left(\dfrac{5}{3}\mathcal{E}_\mathrm{FG}\right)^{-1} $
                       &  ${\kappa_\mathrm{FG}}/{\kappa} = \dfrac{1}{\kappa}\left(\dfrac{10}{9}\rho \mathcal{E}_\mathrm{FG}\right)^{-1}  $
                       &  $P/P_\mathrm{FG} = P\left(\dfrac{2}{3}\rho \mathcal{E}_\mathrm{FG}\right)^{-1}$
%                       &  $\Gamma$
                       \\
\hline\hline \rule[-0.5cm]{0.cm}{1.2cm}
%\begin{tabular}{c}
% Definition
%% (in unit of $X_\mathrm{FG}$)
%\end{tabular}
%                              & $\xi$
%                              & $\dfrac{E}{N} + \rho \left.\dfrac{\partial E/N}{\partial\rho }\right|_N $
 %                             & $2\rho ^2\left.\dfrac{\partial E/N}{\partial\rho }\right|_N
 %                                + \rho ^3 \left.\dfrac{\partial^2 E/N}{\partial \rho ^2}\right|_N$
 %                             & $\rho ^2\left.\dfrac{\partial E/N}{\partial\rho }\right|_N $
% %                             & $2 +
% %                           \rho \left.\dfrac{\partial^2 E/N}{\partial \rho ^2}\right|_N
% %                             \left(\left.\dfrac{\partial E/N}{\rho }\right|_N\right)^{-1}$
%                        \\ \hline \rule[-0.2cm]{0cm}{0.9cm}
\begin{tabular}{c}
 %\medskip
 ~~~ As function of $\xi$ ~~~
 %\medskip
 %(in unit of $X_\mathrm{FG}$)
\end{tabular}
                              & $\xi$
                              & $\xi + \dfrac{k_F}{5}\dfrac{\partial\xi}{\partial k_F}$
                              & $\xi + \dfrac{4}{5}k_F \dfrac{\partial \xi}{\partial k_F}
                                     + \dfrac{k_F^2}{10} \dfrac{\partial^2 \xi}{\partial k_F^2}$
                              & $\xi + \dfrac{k_F}{2}\dfrac{\partial \xi}{\partial k_F}$
%                              & $\begin{array}{c}
%                                \dfrac{5}{3}\left(\xi + \dfrac{4}{5}k_F \dfrac{\partial \xi}{\partial k_F}
%                                     + \dfrac{k_F^2}{10} \dfrac{\partial^2 \xi}{\partial k_F^2}\right) \\
%                                   \times\left(\xi + \dfrac{k_F}{2}\dfrac{\partial \xi}{\partial k_F}\right)^{-1}
%                                  \end{array} $
             \\ \hline\hline \rule[-0.2cm]{0.cm}{0.9cm}
\begin{tabular}{c}
 Close to unitary limit \\
 $-(a_sk_F)^{-1} \to 0$\\
 $(r_ek_F) = 0$
\end{tabular}
                              & $\begin{array}{l}
                                  \xi_0  - (a_sk_F)^{-1}\zeta   \\
                                  - \dfrac{5}{3} (a_sk_F)^{-2}\nu
                                 \end{array} $
                              & $ \begin{array}{l}
                                   \xi_0 - \dfrac{4\zeta}{5}(a_sk_F)^{-1} \\
                                   - (a_sk_F)^{-2}\nu
                                  \end{array}$
                              & $ \begin{array}{l}
                                   \xi_0 - \dfrac{2\zeta}{5}(a_sk_F)^{-1} \\
                                   + \dfrac{2\nu}{3}(a_sk_F)^{-2}
                                  \end{array}$
                              & $ \xi_0 - \dfrac{\zeta}{2}(a_sk_F)^{-1}$
 %                             & $\begin{array}{l}
 %                            \dfrac{5}{3}\left(1 + \dfrac{\zeta}{10\xi_0}(a_sk_F)^{-1} \right. \\
 %                             \left. + \left[\dfrac{2\nu}{3\xi_0} - \dfrac{\zeta^2}{20\xi_0^2} \right](a_sk_F)^{-2}\right)
 %                                 \end{array}$
                         \\ \hline \rule[-0.2cm]{0cm}{0.9cm}
\begin{tabular}{c}
 Unitary limit \\
 $-(a_sk_F)^{-1} = 0$\\
 $ (r_ek_F) \to 0$
\end{tabular}
                              & $\begin{array}{l}
                                  \xi_0 + (r_ek_F)\eta_e \\
                                  + (r_ek_F)^2\delta_e
                                 \end{array} $
                              & $ \begin{array}{l}
                                   \xi_0 + \dfrac{6}{5}(r_ek_F)\eta_e \\
                                         + \dfrac{7}{5}(r_ek_F)^2\delta_e
                                  \end{array}$
                              & $ \begin{array}{l}
                                   \xi_0 + \dfrac{9}{5}(r_ek_F)\eta_e \\
                                   + \dfrac{14}{5}(r_ek_F)^2\delta_e
                                  \end{array}$
                              & $ \begin{array}{l}
                                   \xi_0 + \dfrac{3}{2}(r_ek_F)\eta_e \\
                                   + 2(r_ek_F)^2\delta_e
                                  \end{array}$
%                              & $\begin{array}{l}
%                             \dfrac{5}{3}\left(1 + \dfrac{3}{10}\dfrac{\eta_e}{\xi_0}(r_ek_F) \right. \\
%                              \left. + \left[\dfrac{4}{5}\dfrac{\delta_e}{\xi_0} - \dfrac{9}{20}\dfrac{\eta_e^2}{\xi_0^2}\right](r_ek_F)^2 \right)
%                                  \end{array}$
                              \\
\hline \hline
\end{tabular}
%\end{turn}
  \caption{\label{tab:thermo-limits} Summary of some useful relations for the ground state thermodynamical quantities. Top:
  expression of different quantities as a function of $\xi$, defined in Eq. (\ref{eq:func1}) and its derivatives with respect to $k_F$. Middle: Taylor expansion of the functional in powers of $(a_s k_F)^{-1}$ [$r_e=0$ case], the parameters $\zeta$ and $\nu$ are linked
  to the functional parameters through $\zeta =(9\pi/10)(1-\xi_0)^2 \simeq 1.101$ and $\nu=(3/5)(9\pi/10)^2(1-\xi_0)^3 \simeq 1.165$
  that are both close to $1$ as found empirically in QMC calculations \cite{Cha04,Ast04}. Bottom: Taylor expansion of Eq. (\ref{eq:func1}) in powers of $(r_e k_F)$ at unitarity.}
\end{table*}
%\end{widetext}

%\begin{widetext}
\begin{figure*}
 \includegraphics[scale = 0.55]{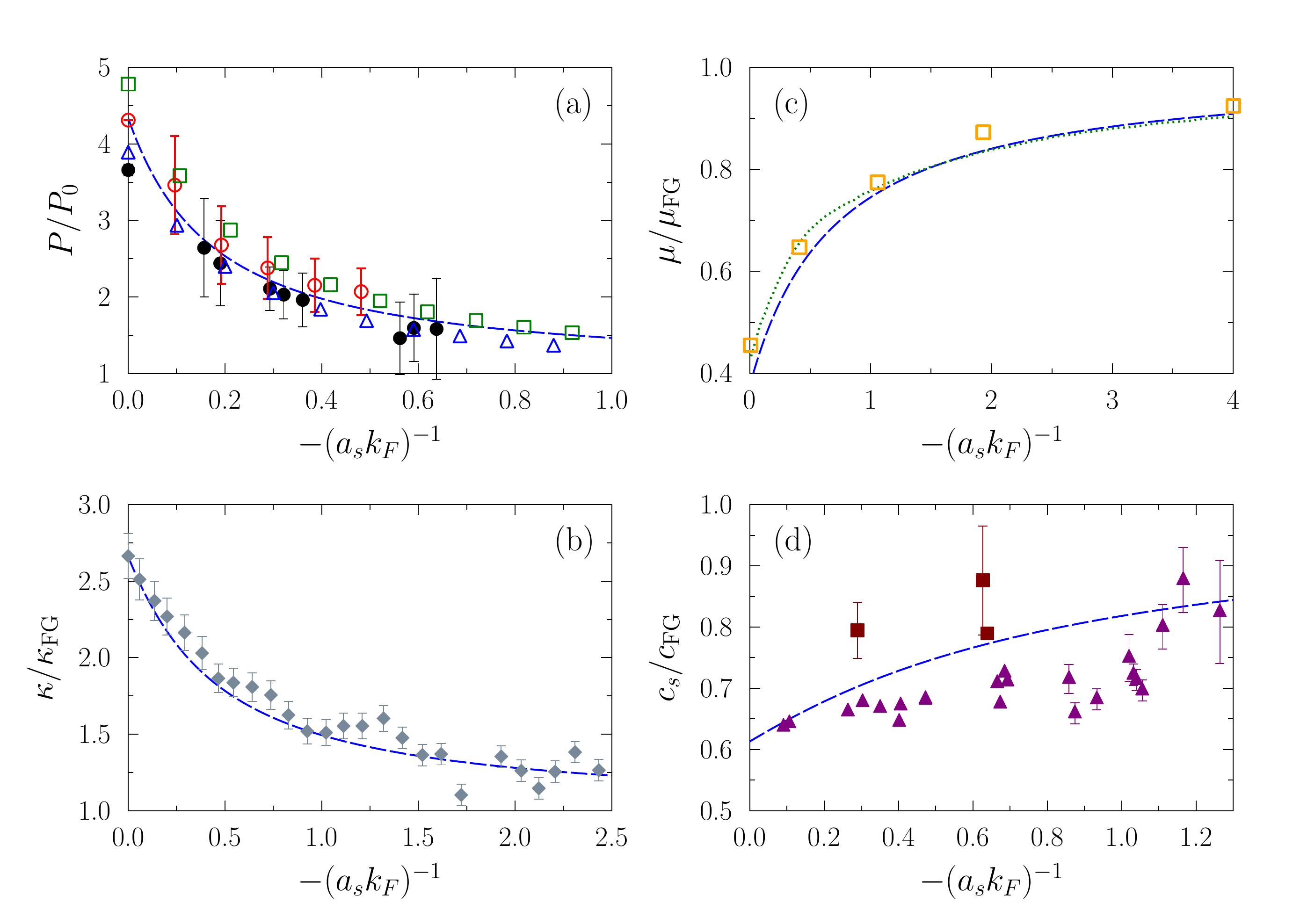}
 \caption{
 The pressure (a), the isothermal compressibility (b), the chemical potential (c) and
 the sound velocity (d) as function of $-(a_sk_F)^{-1}$ obtained from the functional
 \eqref{eq:func1} assuming $r_e = 0$ (blue long dashed line).
 The quantity $P_0$ is defined as $P_0 = (\mu/\mu_{\rm FG})^{5/3}P_{\rm FG} $.
 For comparison, different experimental data or theoretical estimates are shown: (a) black filled circles are data from \cite{Nav10},
 QMC calculations from \cite{Bul08} (red open circles), results obtained from a Nozi\'ere-Schmitt-Rink
 approximation (blue open triangles) \cite{Hu06} or with Green-function method (De Dominicis-Martin
 formalism) (green open squares) \cite{Hau07};
 (b) gray filled diamonds from \cite{Hor16};
 (c) calculation from \cite{Pie05} (open orange squares),
 and the green dotted line is obtained from the best fit to
 the QMC calculations \cite{Ast04} (data taken from figure 3 of \cite{Hu06});
 (d) brown filled squares and  purple triangles are data respectively from \cite{Wei15}
and  \cite{Jos07}.}
 \label{fig:thermo-data}
\end{figure*}

\subsection{Ground state thermodynamical properties}

Thermodynamical properties of atomic gases have been extensively studied both at zero and finite temperature \cite{Zwe11}.
In the present work, we concentrate on the zero-temperature limit. Starting from a density functional approach, we summarize below the expression of selected quantities as a function of derivatives of the energy in homogeneous systems:
\begin{itemize}
 \item {\bf Pressure:}
\begin{eqnarray}\label{eq:pressure-funct}
 P = \left. {\rho ^2}\frac{\partial E/N}{\partial \rho }\right|_N .
 \end{eqnarray}
\item {\bf  Compressibility:}
\begin{eqnarray}
  \kappa = \frac{1}{\rho }\left(\left.\frac{\partial P}{\partial \rho }\right|_N\right)^{-1} ,
\end{eqnarray}
leading to
\begin{eqnarray}
\frac{1}{\rho  \kappa} = \frac{2P}{\rho } + \rho ^2 \left.\frac{\partial^2 E/N}{\partial \rho ^2}\right|_N  . \label{eq:compress-funct}
\end{eqnarray}
\item {\bf Chemical potential:}
\begin{eqnarray}
  \mu &=& \left.\frac{\partial E}{\partial N}\right|_V = \left.\frac{\partial \rho E/N}{\partial \rho }\right|_V =  \frac{E}{N} + \rho  \left.\frac{\partial E/N}{\partial \rho }\right|_V
  \nonumber \\
  &=&  \frac{E}{N} + \frac{P}{\rho }. \label{eq:chem}
%  ~~~~~\Longrightarrow~~~~~
%  \frac{\mu}{\mu_\mathrm{FG}} = \frac{3}{5}\xi + \frac{2}{5}\frac{P}{P_\mathrm{FG}}
 %  = \xi + \frac{k_F}{5}\frac{\partial\xi}{\partial k_F}
\end{eqnarray}
\item {\bf Sound velocity and adiabatic index:} From the above quantities one can deduce the sound velocity $c_s^2 = (m\rho \kappa)^{-1}$ and the adiabatic index  \begin{eqnarray}
  \Gamma = \left.\frac{\rho }{P}\frac{\partial P}{\partial \rho }\right|_N
  = \frac{1}{\kappa P} . %=  \frac{5}{3} \frac{\kappa_\mathrm{FG}P_\mathrm{FG}}{\kappa P}
\end{eqnarray}
\end{itemize}
Alternative expressions can be obtained using directly the quantity $\xi$ introduced in Eq. (\ref{eq:func1}) and its derivatives with respect to the Fermi momentum\footnote{The derivatives with respect to $\rho $ can be transformed
into derivatives with respect to $k_F$ using $3\pi^2\rho  = k_F^3$:
\begin{eqnarray*}
  \frac{\partial }{\partial \rho } = \frac{dk_F}{d \rho }\frac{\partial }{\partial k_F}
    = \frac{\pi^2}{k_F^2}\frac{\partial }{\partial k_F} .
\end{eqnarray*}
}.
These expressions are listed in
table  \ref{tab:thermo-limits} together with specific limits obtained  at or close to unitarity.

In the following, we will normalize these thermodynamical quantities to their corresponding
values for the free FG case, given by:
\begin{eqnarray}
 P_\mathrm{FG}     & \equiv& \frac{2}{3}\rho \mathcal{E}_\mathrm{FG},
 ~~~
 \frac{1}{\kappa_\mathrm{FG}}
                    \equiv \frac{10}{9}\rho \mathcal{E}_\mathrm{FG}, \nonumber \\
 c_\mathrm{FG}^2    &  \equiv& \frac{10}{9m}\mathcal{E}_\mathrm{FG},
 ~~~
 \mu_\mathrm{FG}    \equiv  \frac{5}{3}\mathcal{E}_\mathrm{FG} = \frac{\hbar^2k_F^2}{2m} ,\nonumber
\end{eqnarray}
where $\mathcal{E}_\mathrm{FG}$ is defined in equation \eqref{eq:EFG-def}.

%\begin{widetext}

\begin{figure}
 \includegraphics[scale=0.5]{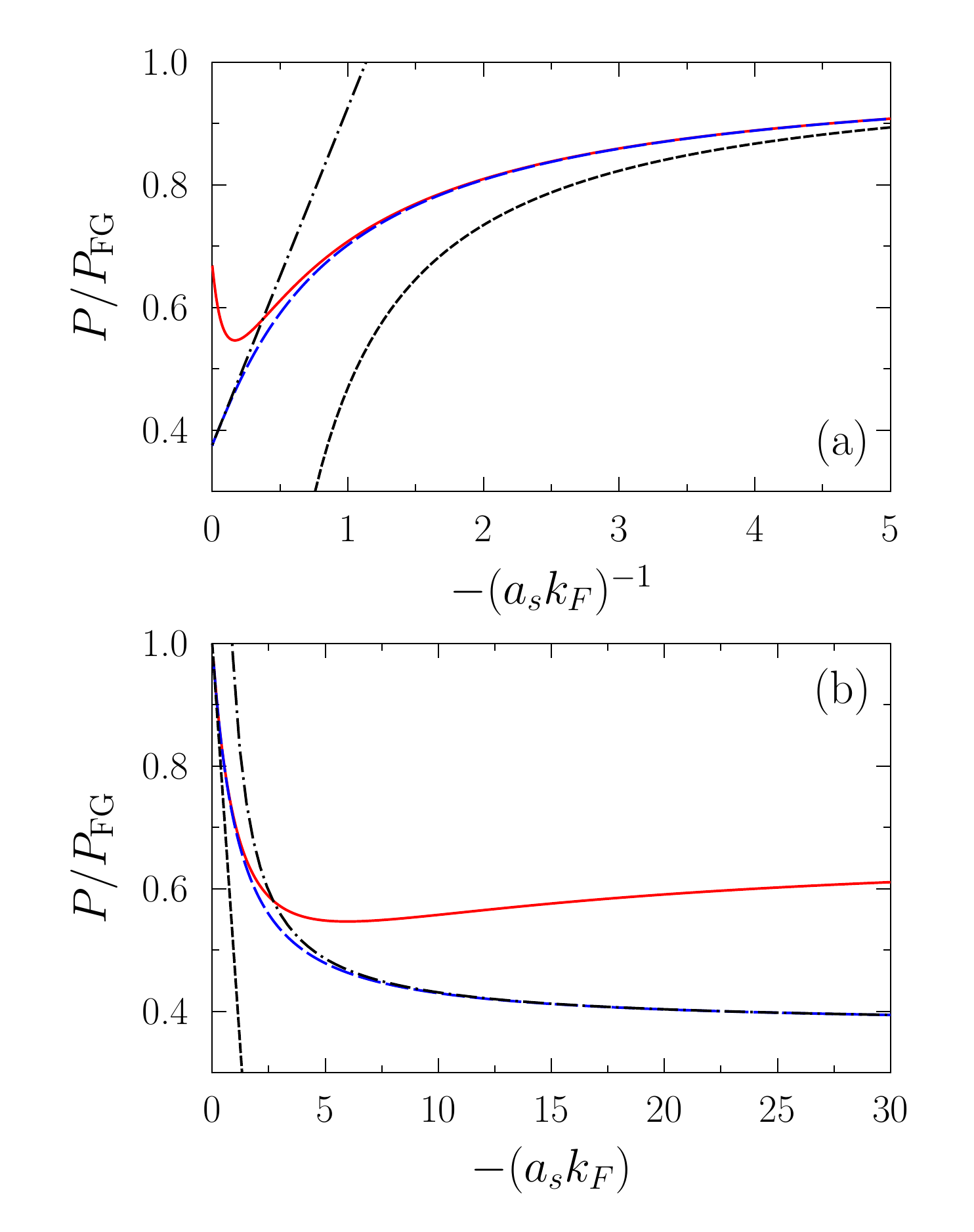}
 \caption{Pressure as a function of $-(a_sk_F)^{-1}$ (a)  or as a function of $-(a_sk_F)$ (b)
 obtained from the functional (\ref{eq:func1}) assuming $r_e=0$ fm (blue long dashed line).
 For comparison, the black short dashed and the black short dot-dashed lines correspond to the Taylor expansion
 to first order in $(a_s k_F)$ or to first order in $(a_sk_F)^{-1}$ respectively.
 The red solid line corresponds to the neutron matter case assuming $r_e = 2.716$ fm. In both cases, $a_s=-18.9$ fm.
 }
 \label{fig:pressure-4fig}
\end{figure}

%\end{widetext}
Thermodynamical quantities obtained using the simple functional for $r_e=0$ and arbitrary large negative
scattering length are compared in Fig. \ref{fig:thermo-data} with various experimental
observations and/or theoretical estimates.  Not surprisingly, the functional reproduces the unitary limit since it has explicitly
been adjusted to reproduce the Bertsch parameter $\xi_0$. We see that, despite its simplicity and the fact that the functional only depends on $\xi_0$,
it is able to reproduce rather well the thermodynamics of Fermi--gas away from unitarity.  It should be noted that none of the Taylor expansions in $(a_s k_F)$ or $(a_s k_F)^{-1}$ would be able to reproduce these quantities from very low to very high densities, as illustrated in Fig. \ref{fig:pressure-4fig} for the pressure. Similar behavior is obtained for the other
quantities shown in Fig.  \ref{fig:thermo-data}.

\subsection{Non-zero effective range effect}

\begin{figure}
 \includegraphics[scale=0.55]{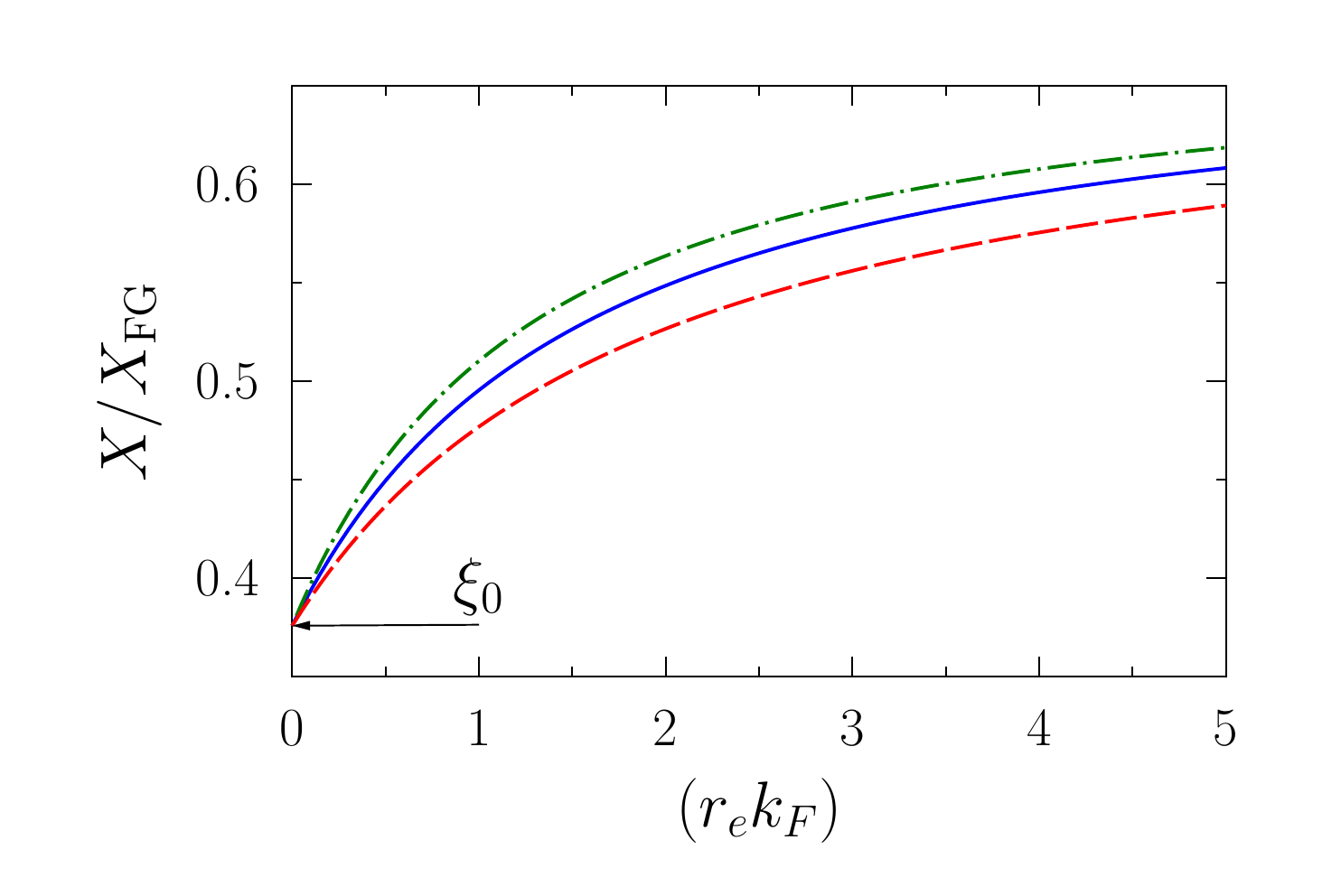}
 \caption{ The pressure [$X=P$] (blue solid line), chemical potential [$X=\mu$] (red dashed line) and inverse compressibility [$X=1/\kappa$]
 (green dot-dashed line) obtained at unitarity using the functional (\ref{eq:funcre}) as a function of $(r_e k_F)$.
 The arrow indicates the unitary limit for $r_e=0$.
 }\label{fig:rekf-dependance}
\end{figure}

In Fig. \ref{fig:pressure-4fig}, we also show an example of evolution of thermodynamical quantities for non-zero effective range relevant for neutron matter.
In the present functional, the effect of $r_e$ is mainly visible at large $(a_s k_F)$ values, i.e. close to unitarity.   To illustrate the dependence
with $r_e$, we consider the strict unitary limit. In that case, the $\xi$ function depends only on $(r_e k_F)$ and we deduce:
\begin{eqnarray}
\xi(a_s k_F \rightarrow - \infty, r_ek_F)   &=& \xi_0 + \frac{\eta_e^2 (r_ek_F)}{\eta_e -\delta_e(r_ek_F)},  \label{eq:funcre}
\label{eq:funcunit_re}
\end{eqnarray}
from which all thermodynamical quantities can be calculated.  The effect of the effective range is predicted to increase
the apparent Bertsch parameter leading also to an increase of the thermodynamical quantities at unitarity.
This is illustrated in Fig. \ref{fig:rekf-dependance} for  the pressure, the chemical potential and the inverse of the compressibility.
We see in particular that the maximal value of $\xi$ at unitarity is $\xi_0 - \eta_e^2/\delta_e =0.66 $ which is almost twice the value
of $\xi_0$ and therefore might be significant.

\subsection{Ground state thermodynamics of neutron matter}

\begin{figure*}
 \includegraphics[scale = 0.55]{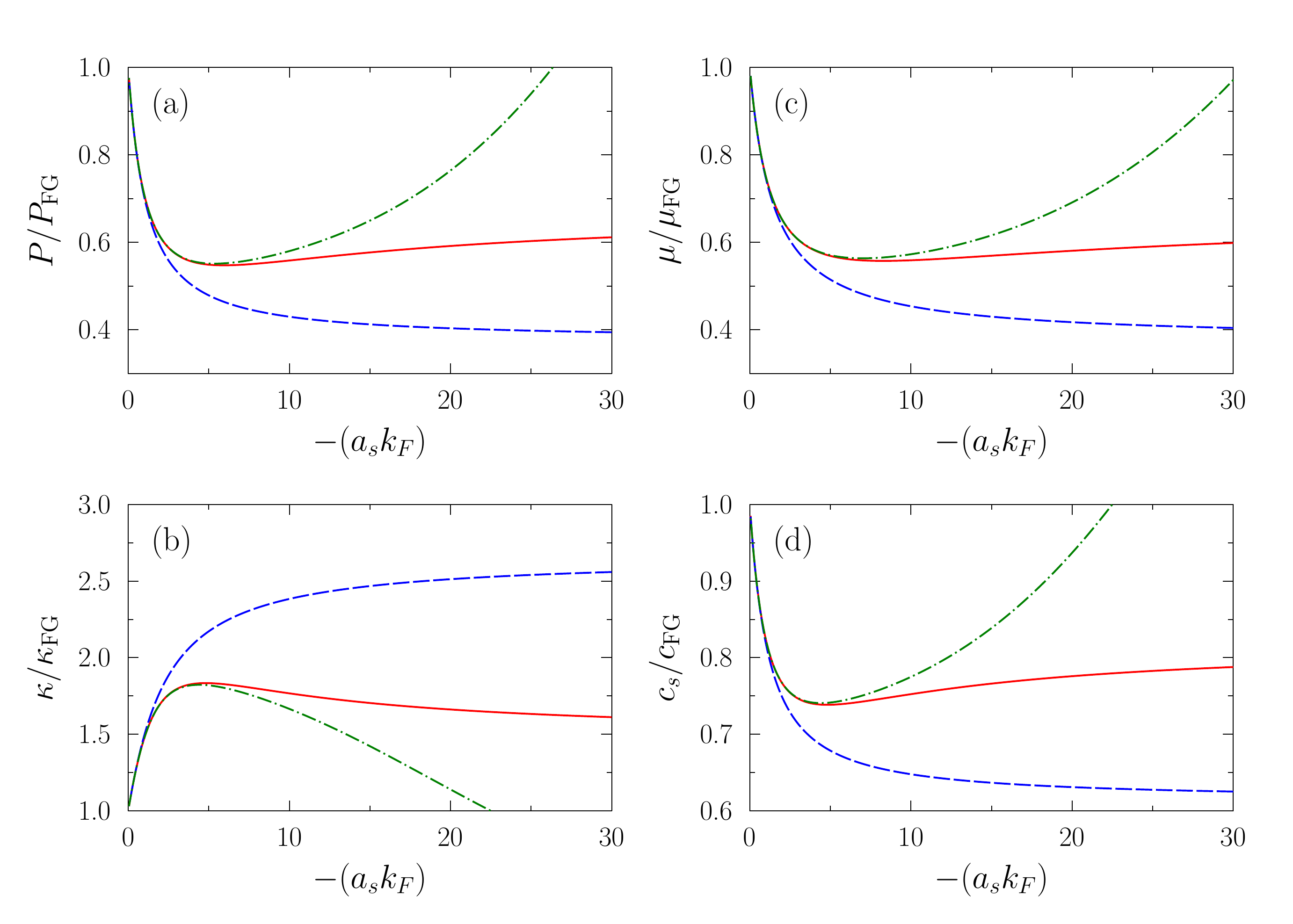}
 \caption{
 The pressure (a), the inverse of the compressibility (b), the chemical potential (c) and
 the sound velocity (d) obtained in neutron matter
as function of $-(a_sk_F)$ using the functional
 \eqref{eq:func1}  (red solid line). The case with $r_e=0$ (cold atom reference) is displayed with blue dashed line.
 The green dot-dashed line corresponds to the result obtain by adding to the functional the leading order contribution
 of the $p$-wave (see text). In this figure, we use $a_s = 18.9$ fm and $r_e= 2.716$ fm so that
 $-(a_sk_F)=10$, $20$ and $30$ correspond respectively to $\rho=0.005$, $0.040$ and $0.135$ fm$^{-3}$.}
 \label{fig:thermo-NM}
\end{figure*}

We have extended the above study to the case of neutron matter for which we anticipate important influence of the effective range $r_e$
as  well as eventually of higher order channels contributions when the density increases. The different thermodynamical quantities obtained using the functional
(\ref{eq:func1}) with realistic values of the low energy constant $a_s$ and $r_e$ are shown in Fig. \ref{fig:thermo-NM}. While at very low density the different quantities are
only slightly affected by effective range, we indeed observe at densities of interest in the nuclear context, i.e. $\rho\simeq 0.05-0.15$ fm$^{-3}$,
differences with the cold atom case.

%\begin{widetext}
%\end{widetext}

\section{Static response in Fermi liquids}

Some of the ground state quantities discussed above are directly connected to the static response of the Fermi system to an external field.
In general, the static response provides interesting insight about the complex internal
reorganization in strongly interacting Fermi liquids \cite{Pin66,Kit05}.
The static or dynamical responses have been the subject of extensive studies in the context of nuclear density functional theory
\cite{Gar92,Pas12,Pas12-b,Pas12-c,Pas15}  especially those based on the Skyrme EDF. As we will see, in the latter
case, the static response strongly depends on the set of parameters used in the Skyrme EDF. Recently, the ab-initio static response of neutron matter
has been obtained using AFDMC for the first time in Refs. \cite{Bur16,Bur17} giving strong
constraints on nuclear EDF.  One surprising result is that the static response is very close to the
free FG response. Below we make a detailed discussion on the static response obtained using the
functional
(\ref{eq:func1}). Since the methodology to obtain the static response is already well documented \cite{Pin66,Kit05}, we only give  the important equations used thorough the article.

\subsection{Generalities on static response}

Let us consider a system described by a many-body Hamiltonian $\hat H$. A static external one-body field, denoted
by $V_{\rm ext}$ is applied to the system leading to a change of its properties. The static response, denoted by $\chi$ contains
the information on how the one-body density and total energy vary with the external field. $\chi$ is defined through:
\begin{eqnarray}
\delta \rho({\mathbf r}) &=& \int d^3 {\mathbf r}' {\chi} ({\mathbf r} - {\mathbf r}') V_{\rm ext} ({\mathbf r}') \label{eq:rhoV},
\end{eqnarray}
where $\delta  \rho({\mathbf r}) =  \rho({\mathbf r}) - \rho_0$,  $\rho({\mathbf r})$ and $\rho_0$ being respectively the local one-body density and
the equilibrium density of the uniform system in its ground state.
From this, one can express the static response formally as
\begin{eqnarray}
\chi({\mathbf r} - {\mathbf r}') & = & \left( \frac{\delta \rho({\mathbf r})}{\delta V_{\rm ext} ({\mathbf r}')} \right).
\end{eqnarray}
Performing the Fourier transform of Eq. (\ref{eq:rhoV}), we simply have:
\begin{eqnarray}
\delta \rho({\mathbf q}) &=&  {\chi} ({\mathbf q}) V_{\rm ext} ({\mathbf q}).
\end{eqnarray}
Following Refs. \cite{Bur16,Bur17}, we  assume $V_{\rm ext} ({\mathbf r}) =2  \sum_{\mathbf q} v_q
 \cos({\mathbf q} \!\cdot\! {\mathbf r})$. The Fourier transform of $V_{\rm ext}$
at ${\mathbf q}$ is then simply a constant and we have for the energy:
\begin{eqnarray}
E({\mathbf q}) & = & E_0 + \frac{\chi (q) }{\rho_0} v^2_q + \cdots
\end{eqnarray}

\begin{figure}[htbp]
\includegraphics[width=0.8\linewidth]{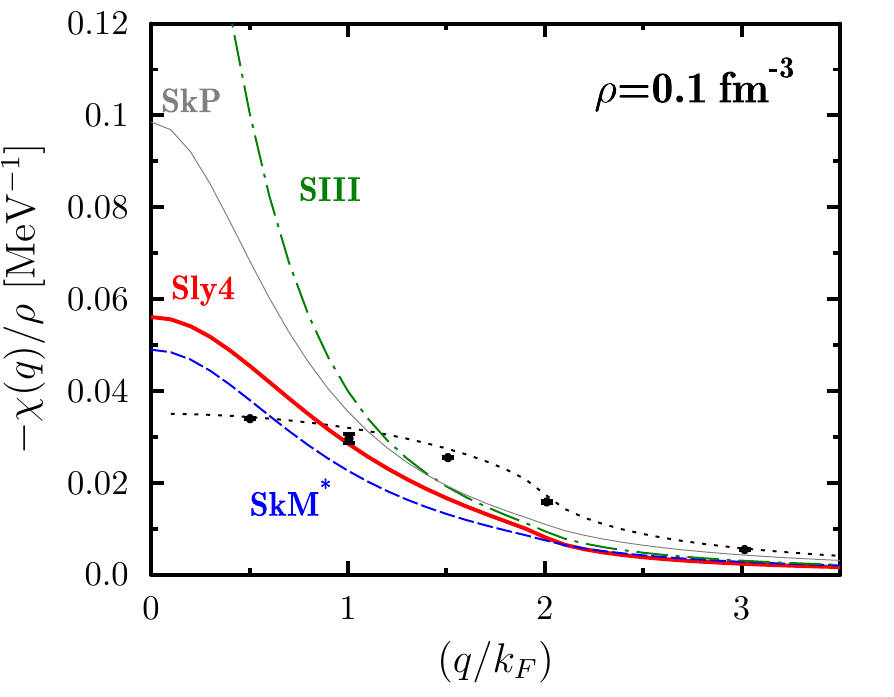}
\vspace{-5mm}
\caption{Static response of neutron matter at density $\rho = 0.1$ fm$^{-3}$ obtained with the Skyrme EDF using the Sly4 (red thick solid line),
SkM$^*$ (blue dashed line), SkP (gray thin solid line) and SIII (green dot-dashed line)
sets of Skyrme parameters (see \cite{Mey03} for their values). Note that the EDF results are obtained by neglecting the spin--orbit term in the functional.
Adding the spin--orbit contribution does not change significantly the result.
The black circles (with errorbars) are those
of Ref. \cite{Bur17} while the black dotted  line is the free Fermi--gas static response. }
\label{fig:skfull01}
\end{figure}

The static response function is directly linked to the compressibility $\kappa$ discussed above due to the asymptotic relationship:
\begin{eqnarray}
 \lim_{q\to0} \chi(q) = - \rho ^2 \kappa. \label{eq:chikappa}
\end{eqnarray}
The function $\chi(q)$ or its dynamical equivalent has been extensively studied for the Skyrme EDF \cite{Pas15}.  In Fig. \ref{fig:skfull01}, we give examples
of results obtained using different sets of Skyrme parameters and compared them with the AFDMC
results of Refs. \cite{Bur16,Bur17}. It is
clear from this figure that there is a large dispersion in the Skyrme EDF response depending on the parameter sets. Such a dispersion is not surprising since it is well known
that the neutron equation of state is weakly constrained in Skyrme EDF (see for instance \cite{Bro00}). In all cases, even if the Skyrme EDF gives reasonable neutron matter EOS, significant difference is observed with the exact AFDMC result for all ranges of $q/k_F$. This figure also illustrates the fact that the exact
result  is very close to the free Fermi--gas response.

\begin{figure}[htbp]
\includegraphics[width=0.8\linewidth]{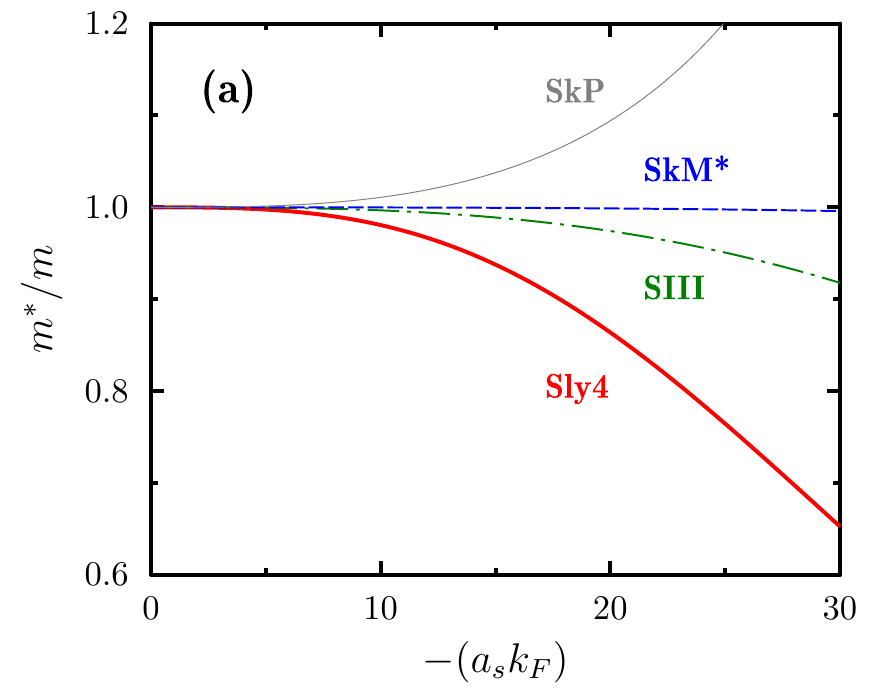} \\
\includegraphics[width=0.8\linewidth]{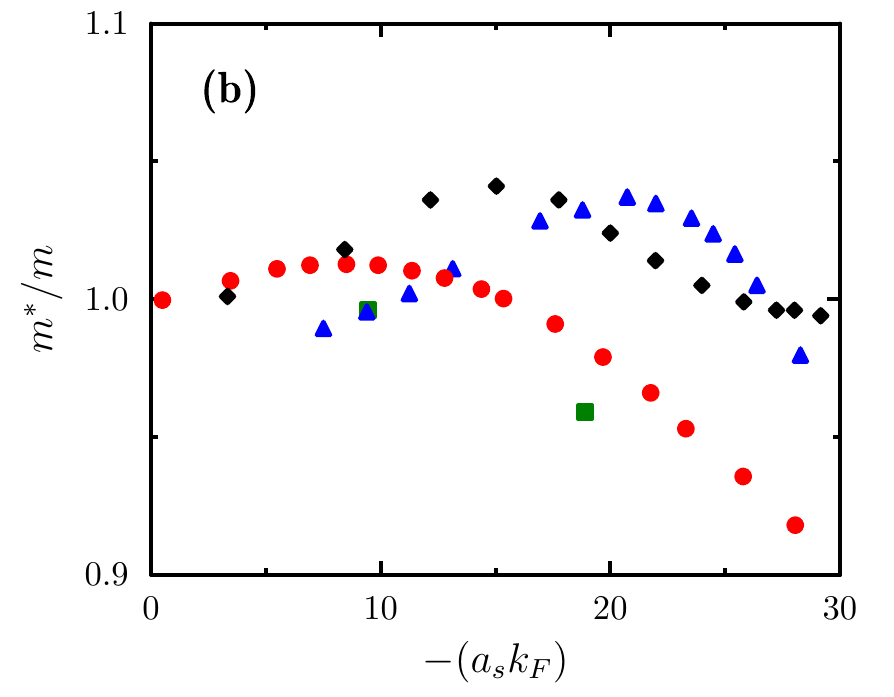}
\vspace{-5mm}
\caption{ (a) Evolution of the effective mass $m^*/m$ in neutron matter as a function of $-(a_s k_F)$ (with $a_s=-18.9$ fm) using
 Skyrme EDF for the set of parameters used in Fig. \ref{fig:skfull01} with the same lines convention. (b) Evolution of the effective mass
 in neutron matter deduced from
selected many-body calculations: blue triangles  \cite{Sch03}, red circles \cite{Wam93}, green squares \cite{Fri81}, black diamonds \cite{Dri14}.}
\label{fig:emassskyrme}
\end{figure}

An important ingredient of the response in Skyrme EDF is the evolution of the effective mass as a function of the density. Such evolution is shown
in Fig. \ref{fig:emassskyrme}(a). Again, large differences are observed between the different sets
of Skyrme EDF. We also show for comparison,
the effective mass obtained in neutron matter using alternative many-body techniques.  In these cases, the deduced effective mass are closer
to the bare mass but still significantly differ with each other.

In the absence of clear guidance for the effective mass behavior, we simply assume below that $m^*/m=1$.  Then, the calculation of the
static response within the density functional approach reduces to
\begin{eqnarray}
\chi(q) & = &\frac{ \chi_0(q)}{1 - G(\rho ) \chi_0(q)},
\end{eqnarray}
where $ \chi_0(q)$  is the response of the free gas given in term of the Lindhard function (dashed line in Fig. \ref{fig:skfull01})
\begin{eqnarray}
\chi_0 ({ q})&=& - \frac{m k_F}{2\pi^2 \hbar^2} \left [ 1 + \frac{k_F}{q}
\left ( 1 - \left ( \frac{q}{2k_F} \right )^2 \right ) \ln \left | \frac{q + 2k_F}{q - 2k_F} \right |   \right ], \nonumber \\
&\equiv& - \frac{m k_F}{\pi^2 \hbar^2} f(q/k_F).
\label{eq:Lindhard}
\end{eqnarray}
The density-dependent coefficient $G(\rho )$ is obtained from the second derivative of the energy density
after subtraction of the kinetic contribution. Explicitly, the energy can be rewritten as an integral over the energy density through:
\begin{eqnarray}
E & = & \int \left\{ {\cal K} ({\mathbf r} ) + {\cal V} ({\mathbf r} )  \right\}  d^3 {\mathbf r} , \nonumber
\end{eqnarray}
where ${\cal K}$ is the kinetic energy density and ${\cal V}$ is the potential energy density. For uniform system, $G(\rho )$ is given by:
\begin{eqnarray}
G(\rho ) & = &  \left( \frac{\partial^2 {\cal V}}{\partial^2 \rho } \right),
\end{eqnarray}
that could again eventually be transformed as partial derivatives with respect to $k_F$.

\subsubsection{Static response in unitary gas}

\begin{figure}[htbp]
\includegraphics[width=0.9\linewidth]{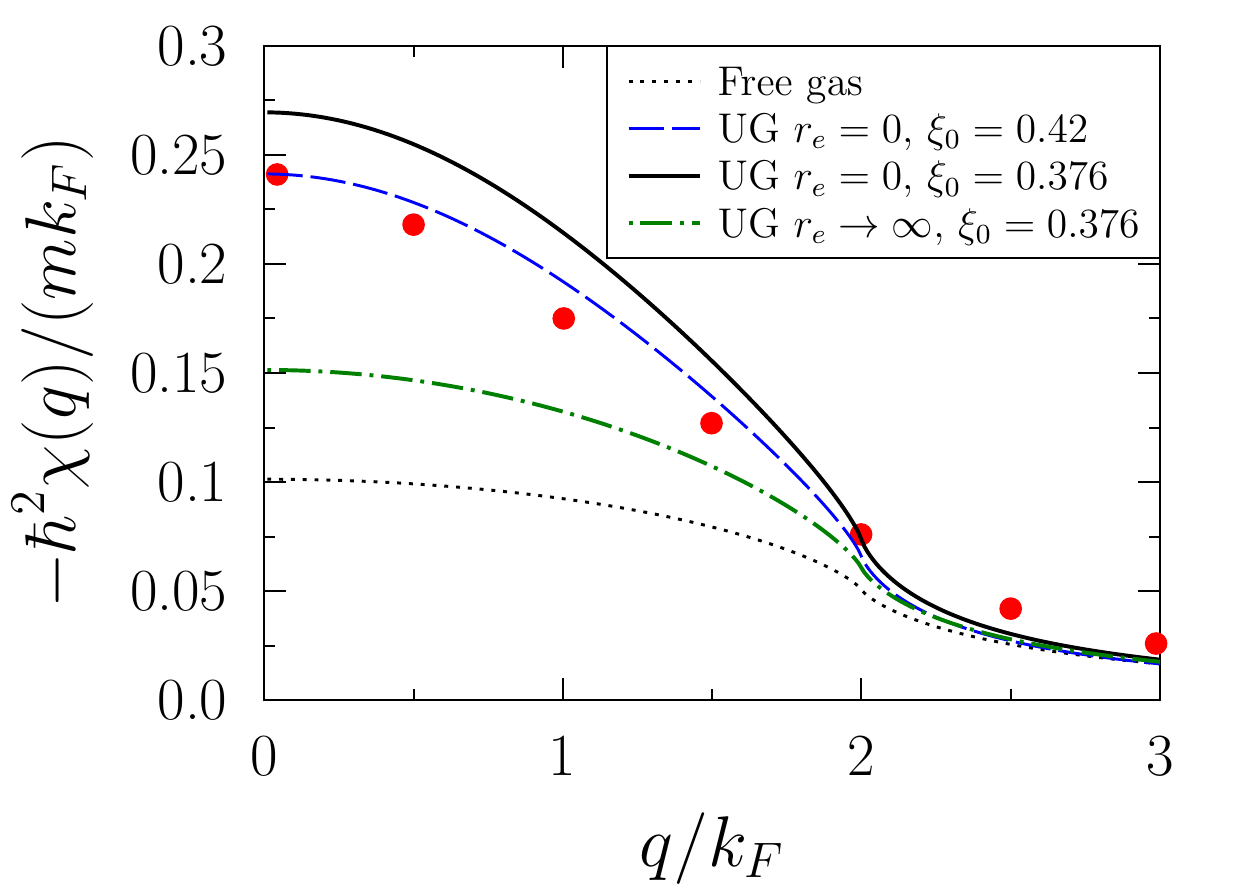}
\vspace{-5mm}
\caption{ Static response of unitary gas (UG) obtained with the functional (\ref{eq:funcunit_re}) assuming $r_e=0$ (black solid line) or
$r_e \rightarrow +\infty$ (green dot-dashed line). The latter case corresponds to the maximum effect induced by $r_e$ and
corresponds to an effective Bertsch parameter equal to $0.66$.
The free Fermi--gas response (black dotted line) is also shown as a reference. For comparison, the red filled circles
corresponds to the static response at unitarity obtained in ref. \cite{For14} using the SLDA where a slightly different value of the Bertsch parameter
$\xi_0= 0.42$ was assumed. For completeness, we also show (blue dashed line) the UG result obtained with (\ref{eq:funcunit_re}) for
$\xi_0= 0.42$.}
\label{fig:staticUG}
\end{figure}

We consider first the strict unitary gas limit with $r_e=0$. In this case, we have
\begin{eqnarray}
{\cal V}(\rho ) =  \frac{3}{5} \frac{\hbar^2}{2m} (3\pi)^{2/3} \rho ^{5/3}(\xi_0 - 1) . \label{eq:Vxi0}
\end{eqnarray}Using this expression, we obtain:
\begin{eqnarray}
\chi(q) & = & - \frac{m k_F}{\pi^2 \hbar^2}\frac{f(q/k_F)}{ \left[1 + (\xi_0 - 1)f(q/k_F)\right]}
\end{eqnarray}
where $f(q/k_F)$ is defined through Eq. (\ref{eq:Lindhard}). In particular, we see immediately that
$\chi(0) = \chi_0 (0) / \xi_0$, with $\chi_0 (0) =  - \frac{m k_F}{\pi^2 \hbar^2} \displaystyle$
that is a direct consequence of the property (\ref{eq:chikappa}) and of the fact that $\kappa_{\rm FG}/\kappa = \xi_0$
in unitary gas (see table \ref{tab:thermo-limits}).

The full static response is shown in Fig.
\ref{fig:staticUG} and compared to the result obtained in Ref. \cite{For14} using the Superfluid Local Density Approximation (SLDA)
proposed in Refs \cite{Bul07a,Mag09}. The static response calculated with
the functional (\ref{eq:Lindhard}) perfectly matches the SLDA result when the Bertsch parameter is
artificially increased to match the one
used in the SLDA. Note that, the SLDA assumed a significant contribution from the effective mass and account also for superfluid
effects, which is not the case in the present work. Therefore, an indirect conclusion is that the static response for UG
does not seem to be significantly influenced in particular by superfluidity.  We would like to insist on the fact that this is most probably  specific to the response to a static field. Indeed, due to superfluidity, the dynamical response will present a low energy
mode, the so-called Bogoliubov-Anderson mode that has been studied for instance in \cite{Ast14}.  The superfluid nature of
Fermi gas at unitarity has been unambiguously directly probed by the presence of lattice of quantized vortices in Ref. \cite{Zwe05}.

Note finally that the matching of the static response obtained with the two functional approach observed in Fig. \ref{fig:staticUG}
is an interesting information but it does not mean that the static response is the correct one at unitarity. A comparison with an exact calculation would be desirable (see also discussion in section \ref{sec:critical}).

The effect of $r_e$ at unitarity can be studied using the generalization of Eq. (\ref{eq:Vxi0}):
\begin{eqnarray}
{\cal V}(\rho ) = E_{\rm FG}(\rho ) [\xi(r_e k_F)  - 1], \nonumber
\end{eqnarray}
 where $\xi(r_e k_F)$ is given by Eq. (\ref{eq:funcre}). The predicted influence of $r_e$ is illustrated in Fig. \ref{fig:staticUG}.
 The effective range induces a global reduction of $\chi(q)$. The maximum effect is achieved by considering the limit $r_e \rightarrow + \infty$ (green dot dashed curve).

\subsection{Static response in neutron matter}

The neutron matter differs from the unitary gas by a finite value of the scattering length as well as significant effect of the
effective range even at rather low density \cite{Lac17}. When the density increases, it is also anticipated that higher
partial waves of the interaction contribute. Since our aim is to compare with the recent result of ref. \cite{Bur17} where the AV8 interaction has been
used, we use the functional (\ref{eq:func1}) using the AV8 values for the different parameters:
$a_s=-19.295~\mathrm{fm}$ and $r_e=2.716~\mathrm{fm}$. We note that, the functional (\ref{eq:func1}) reproduces
well the energy for rather low densities $\rho < 0.02-0.03$ fm$^{-3}$ \cite{Lac17} while the static
responses of Ref.  \cite{Bur17}
have been obtained for $\rho =0.04$ and $0.1$ fm$^{-3}$ which is not optimal for the comparison.

\begin{figure}
 \includegraphics[scale=0.55]{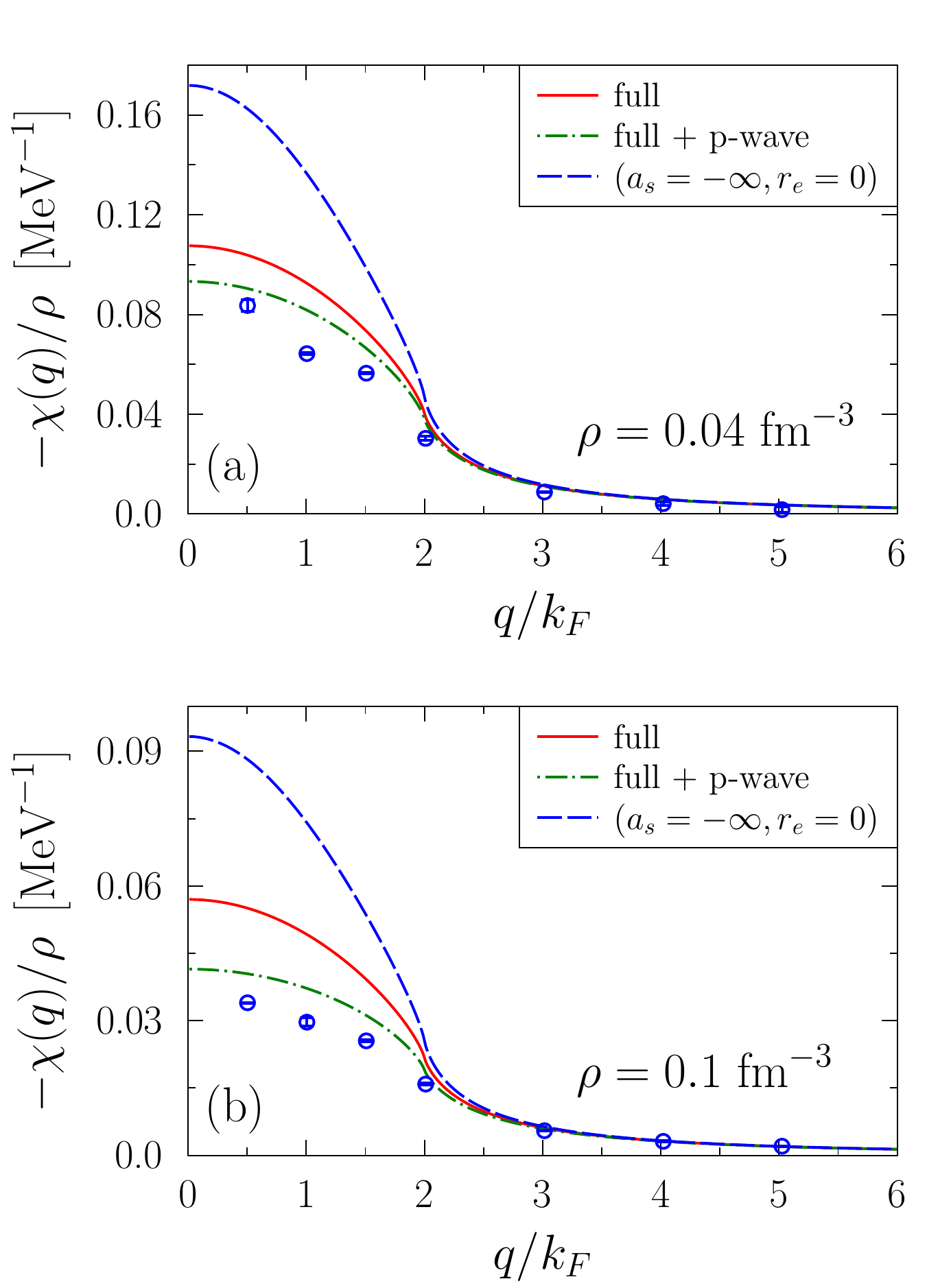}
 \caption{
 Static response function obtained with the functional (\ref{eq:func1})
 as function of $q/k_F$ for (a) $\rho  = 0.04~\mathrm{fm^{-3}}$  and
 (b) $\rho  = 0.1~\mathrm{fm^{-3}}$. The
  blue dashed line and   red solid line corresponds respectively to the UG case  (cold atom reference) and $r_e= 2.716 ~\mathrm{fm}$
  (neutron matter case). Note that at the density considered, the UG case cannot be distinguished from the neutron matter
  assuming $r_e=0$.
  %%%
At both densities, the AFDMC results of Refs. \cite{Bur16,Bur17}
are shown with blue open circles. The green dot-dashed line
 finally display the result obtained by adding the $p$-wave contribution to the functional (\ref{eq:func1}).
 Consistently with the use of the AV8 interaction in the AFDMC calculation, we use a value $a^3_p=0.0916$ fm for the $p$-wave scattering volume.
 %
 %
  %on the bottom (red solid line) compared to three limits:
 %(i) zero range limit $r_e =0$ (purple short dashed line);
 %(ii)  unitary limit $a_s \to -\infty$ (orange doted line);
 %(iii) the limit $a_s \to -\infty ~~\&~~ r_e=0$ (green dot-dashed line).
 %The red open squares and the blue open circles are the results
 %obtained by \cite{Bur16}. Error bars are also reported.
 %In both cases, we use the AV8 scattering parameters: $a_s = -19.295$ fm and $r_e = 2.716$ fm.
% \code{dyn-resp-UF.for}
 }\label{fig:stat-neut}
\end{figure}

We show in Fig. \ref{fig:stat-neut} the static response obtained from the functional
(\ref{eq:func1}) and compare it to the AFDMC results of Refs.
\cite{Bur16,Bur17}. While slightly overestimated, especially in the highest density considered, we first observe that the new
functional is in much better agreement than the empirical functional considered in Fig. \ref{fig:skfull01}.  For the considered densities, as
underlined in Ref. \cite{Lac17}, the functional (\ref{eq:func1}) can be accurately replaced by its unitary gas limit, i.e. taking
$-(a_s k_F)^{-1} = 0$. Indeed, replacing $\xi$ entering in the full functional by $\xi$ given by Eq. (\ref{eq:funcre}) leads almost to the same
total energy and static response (not shown). Still, the static response obtained by neglecting the $r_e$ effect is rather far
from the static response obtained with the physical $r_e$, underlying the key role played by the effective range.

Following Ref. \cite{Lac17}, we also study the possible influence of the $p$-wave contribution by adding simply its leading order contribution
to the energy, that is
\begin{eqnarray}
\frac{E_{p}}{E_{\rm FG} } & = & \frac{1}{\pi} (a_p k_F )^3. \nonumber
\end{eqnarray}
The results are displayed in Fig. \ref{fig:stat-neut}. We see that the $p$-wave term, treated simply by its leading order contribution
does contributes to the static response and, most importantly, the result is very close to the ab-initio one.
For the sake of completeness, we also report in Fig. \ref{fig:thermo-NM} the different thermodynamical quantities obtained by including the $p$-wave term.   However the contribution should only be taken here as indicative.
As noted already in Ref. \cite{Lac17}, the inclusion of the leading term of the $p$-wave, produces a rather large, most probably unphysical, contributions
to the different quantities and when density increases one should a priori properly account for the $p$-wave contribution accounting from
the re-summation of the s-wave effect as illustrated in Ref. \cite{Sch05}. This is out of the scope of the  present work.

%\textcolor{green}{(Denis:  in view of the above result, I would remove the following text in green)
%For the density
%considered, especially for $\rho =0.1$ fm$^{-3}$ we are $(r_e k_F)$ is large and we see that the static response almost match
%the UG limit with $r_e \rightarrow + \infty$.  In this limit, the response is essentially driven by a single parameter that is
%$\xi_0 - \eta_e^2/\delta_e$. While the value of $\xi_0$ is now rather well constrained both experimentally and theoretically,
%values of $\eta_e$ and $\delta_e$ are scarcely known. Concentrating our attention to the lowest density where higher
%contribution effects to the energy are expected to be less important, we see that a value $- \eta_e^2/\delta_e = xxx$ would give a perfect agreement.
%Assuming the same $\eta_e$, this would lead to $\delta_e=- yyy$.  This illustrates the extreme sensitivity
%of the static response to the effective range property.
%}

\begin{figure}
 \includegraphics[scale=0.55]{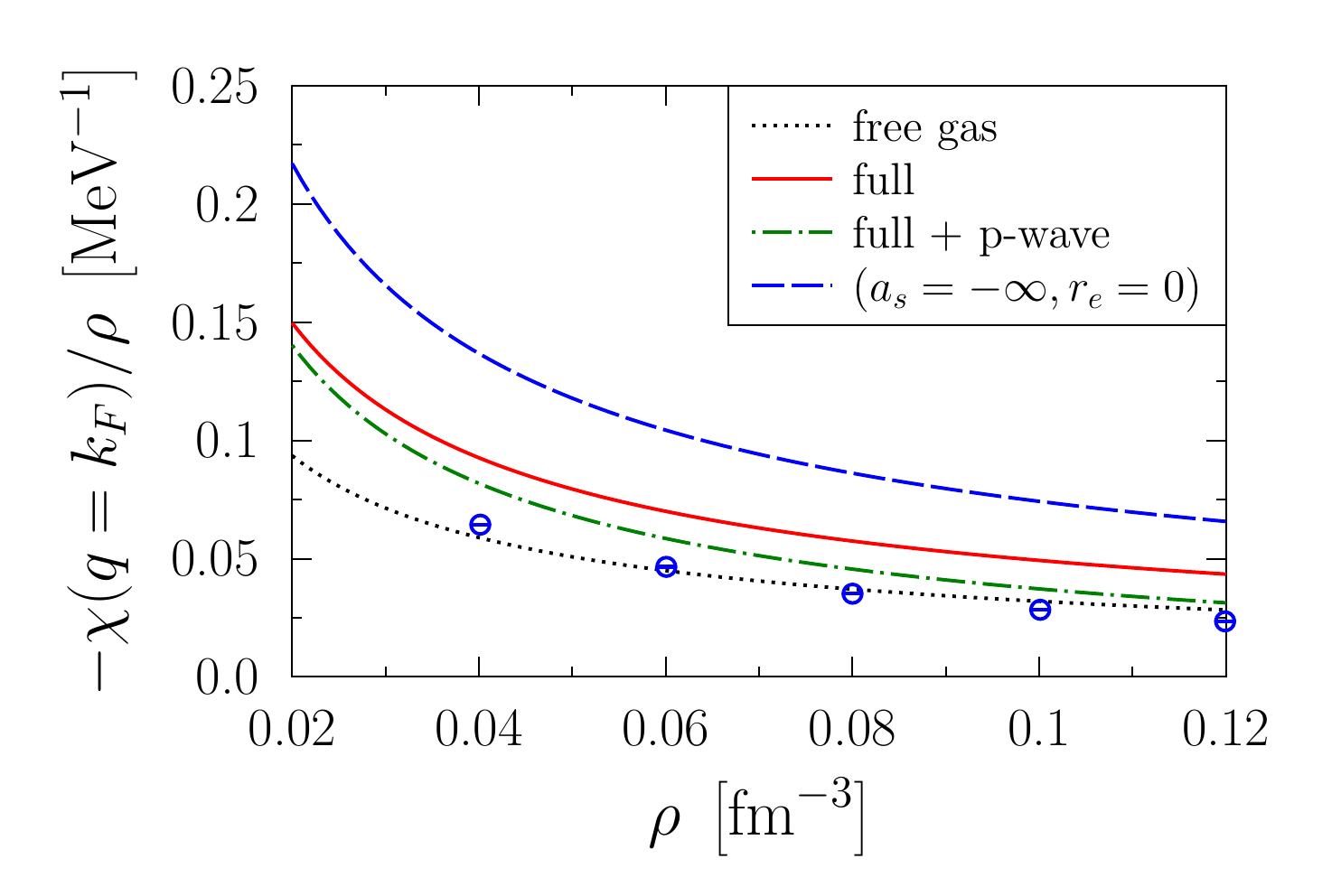}
 \caption{Evolution of the normalized static response as a function of density for $q=k_F$ (same as Fig. 15 of Ref. \cite{Bur17}).
 The  AFDMC results of Ref. \cite{Bur17} is shown by blue open circles.
 The result of the functional (\ref{eq:func1}) with $r_e=2.176$ fm and $r_e=0$ fm are shown by red solid and blue dashed line respectively.
 The result obtained by adding the $p$-wave contribution to the functional is shown by green dot-dashed line. The black dotted line corresponds
 to the free Fermi--gas limit.
 }\label{fig:normresprho}
\end{figure}

Finally, to systematically quantify the effects of finite $a_s$, influence of $r_e$ and $p$-wave, we have reproduced the Fig. 15 of Ref. \cite{Bur17}
where the normalized response function is shown for different densities (Fig. \ref{fig:normresprho}). In this figure, we can clearly see the importance
of the effective range and to a lesser extend, the slightly smaller effect of the $p$-wave. Still the free Fermi--gas case is the one that best reproduces
the AFDMC result.  However, this is most probably accidental in view of the strong interaction at play in nuclear systems.

We finally would like to mention that we are unable with the present density functional to reproduce
the strong increase of the response function as $q \rightarrow 0$ that is observed in the AFDMC.
This limit is directly connected to the compressibility (see Eq. (\ref{eq:chikappa})).
The compressibilities predicted by our EDF are $\rho\kappa = 0.108 ~\mathrm{MeV}$ and $\rho\kappa =
0.057 ~\mathrm{MeV}$ at $\rho = 0.04 ~\mathrm{fm}^{-3}$ and $\rho = 0.1 ~\mathrm{fm}^{-3}$
respectively. These values are lower than those reported in Ref. \cite{Bur17} which are
respectively $\rho\kappa = 0.19 ~\mathrm{MeV}$ and $\rho\kappa = 0.089 ~\mathrm{MeV}$ at $\rho =
0.04 ~\mathrm{fm}^{-3}$ and $\rho = 0.1 ~\mathrm{fm}^{-3}$.

\section{Collective response in the hydrodynamical regime}

We conclude this work by using previous results to study the collective excitations in cold atoms
and neutron matter in the hydrodynamical regime. For boson systems,  the hydrodynamical regime is well
documented \cite{Lip08,Str96}. Similar technique can be applied to fermionic superfluid systems.  Note that here, we
do not include explicitly the pairing correlations through the anomalous density. However, the fact that we properly
describe the total energy of cold atoms, is an indication that pairing effect is accounted for in some way. For superfluid
Fermi system, the hydrodynamical regime is justified  when the collective frequency is below the energy necessary to break a Cooper
pair (see for instance \cite{Bul05}.
Our aim is to study the dynamical response of a system
confined in a trap, describe by an external potential  $U({\mathbf r})$. At equilibrium, the external field is
counterbalanced by the internal pressure leading to the equilibrium equation:
\begin{eqnarray}
\nabla^2 P ({\mathbf r}) &=& - \frac{1}{m} \nabla \left[ \rho_0 ({\mathbf r}). \nabla U({\mathbf r}) \right],
\end{eqnarray}
where $\rho_0$ denotes the equilibrium density while $P$ is the pressure at equilibrium given by Eq. (\ref{eq:pressure-funct}).
We now consider small amplitude oscillations around equilibrium such that $\rho({\mathbf r},t) = \rho_0 ({\mathbf r})+
\delta \rho({\mathbf r},t)$ with $\delta \rho({\mathbf r},t) = \rho_1 e^{i\omega t} + {\rm h.c.}$. The linearization of the hydrodynamical
equation leads to the equation:
\begin{equation}
 \omega^2 \rho_1({\bf r}) =  -\frac{1}{m}\nabla\cdot\left[\rho_1({\bf r}) \nabla U({\bf r}) \right] - \nabla^2\left[c_0^2({\bf r})\rho_1({\bf r}) \right],
\end{equation}
where $c^2_0 ({\bf r})$ is the local sound velocity defined through: $m c^2_0 ({\bf r}) \equiv dP({\mathbf r})/ d \rho_0 ({\mathbf r})$.
This equation has been used in several works to study collective oscillations in Fermi--gas around
unitarity \cite{Bul05,Str96,Hei04,Adh08}.
Below we extend these studies by considering possible effect of non-zero $r_e$ and by going from cold atoms to neutron matter.

\subsection{Adiabatic index in cold atoms and neutron matter}
\begin{figure}
 \includegraphics[scale=0.55]{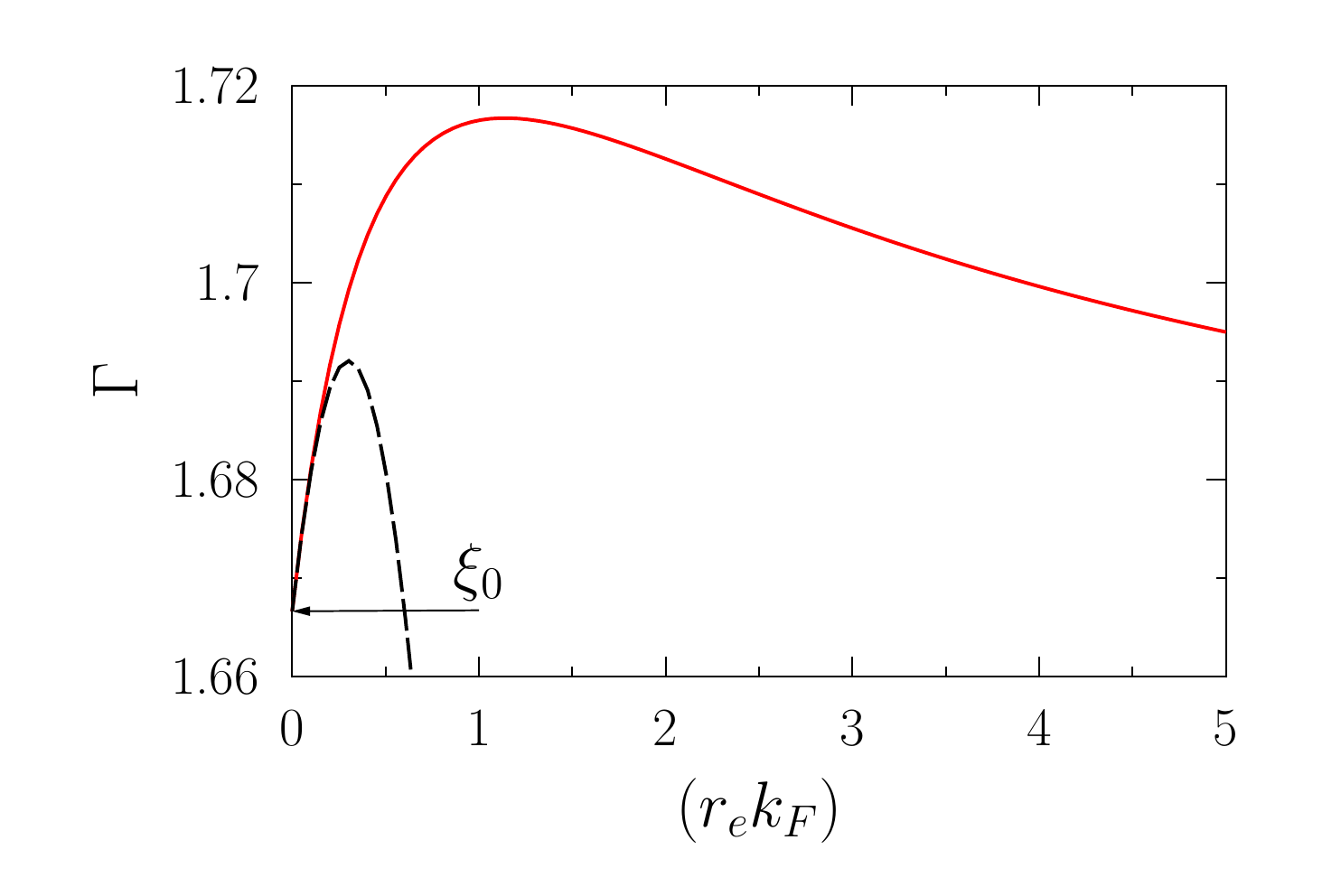}
 \caption{\label{fig:gindexUG} Evolution of the non-adiabatic index $\Gamma$ at unitarity (red solid line)
deduced from the functional (\ref{eq:funcre}) as a function of $(r_e k_F)$.
 The arrow indicates the unitary limit for $r_e=0$.
 For comparison, the black short dashed line corresponds to the Taylor expansion of $\xi$
 to second order in $(r_e k_F)$.}
\end{figure}

For the sake of simplicity, we assume that the system has a polytropic  equation of state, i.e. that we simply have:
\begin{eqnarray}
  P({\bf r}) \propto \rho_0^\Gamma({\bf r}),
\end{eqnarray}
where $\Gamma$ is the adiabatic index in the center of the trapping potential. %that is given by:
%\begin{eqnarray}
%\Gamma  = \left.\frac{\rho_0 ({\bf r})}{P  ({\bf r})}\frac{\partial P  ({\bf r})}{\partial \rho_0
%({\bf r})}\right|_N .
%\end{eqnarray}
As in infinite system, we have the relation $\Gamma = (\kappa_cP_c)^{-1}$
where $P_c$ and $\kappa_c$ denote the pressure and the compressibility in the center of the
trapping potential at equilibrium given by Eqs. (\ref{eq:pressure-funct}) and
(\ref{eq:compress-funct}). The quantity $\Gamma$ has been studied in cold atoms for varying $-(a_s
k_F)$ in Ref. \cite{Cha04}. For vanishing $r_e$, it is know that $\Gamma \rightarrow 5/3$ both in
the unitary limit and in the low density regime. For UG, when $r_e$ could not be neglected anymore,
using the functional
(\ref{eq:funcre}), we predict that $\Gamma$ will deviate from $5/3$. The dependence of $\Gamma$ with
the effective range is shown in Fig. \ref{fig:gindexUG}. We see that $\Gamma$ first increases and
then decreases.
In the extreme limit $r_e \rightarrow +\infty$, it is possible to show that we again obtain $\Gamma \rightarrow 5/3$.

\begin{figure}
 \includegraphics[scale=0.55]{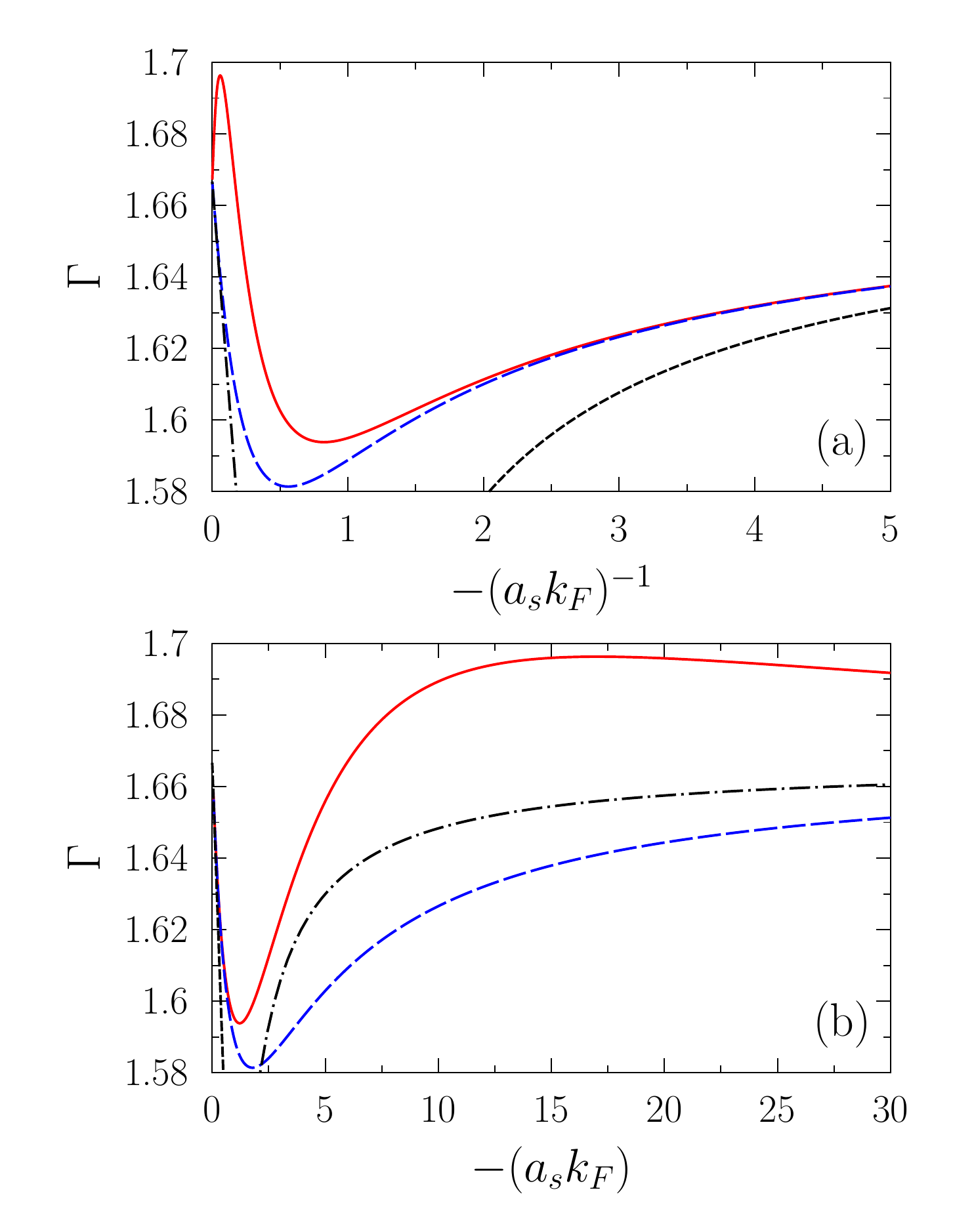}
 \caption{
Adiabatic index as a function of $-(a_sk_F)^{-1}$ (a)  or as a function of $-(a_sk_F)$ (b)
 obtained from the functional (\ref{eq:func1}) assuming $r_e=0$ fm (blue long dashed line).
 For comparison, the black short dashed and the black short dot-dashed lines correspond to the Taylor expansion
 to first order in $(a_s k_F)$ or to first order in $(a_sk_F)^{-1}$ respectively.
 The red solid line corresponds to the neutron matter case assuming $r_e = 2.716$ fm. In both cases, $a_s=-18.9$ fm.
 }
 \label{fig:gindex}
\end{figure}

More generally, we illustrate the dependence of $\Gamma$ obtained with or without effective range effects in Fig. \ref{fig:gindex}
 for low energy constants taken from neutron matter.
For $r_e=0$, we qualitatively and quantitatively reproduce the result of Ref. \cite{Cha04} with the presence of a minimum in $\Gamma$
for $-2.5<(a_s k_F)< 0$. While the minimum persists for non-vanishing $r_e$, we observe that it is slightly shifted to lower values of $|a_s k_F|$.
Overall, we see that $r_e$ significantly affects the evolution of $\Gamma$ that now presents a maximum and approaches $\Gamma=5/3$
from above as $-(a_s k_F) \rightarrow +\infty$.

\subsection{Collective frequencies in anisotropic trap}

As shown in Ref.  \cite{Hei04}, assuming polytropic equation of state leads to rather simple expression
of the collective oscillations in deformed systems. More precisely, we consider here a system confined
in an anisotropic trap
\begin{eqnarray}
 U({\bf r}) = \frac{m}{2} \omega_0^2 \left( x^2 +  y^2 + \lambda^2 z^2 \right),
\end{eqnarray}
where $\lambda$ gives a measure of the anisotropy, with $\lambda < 1$ and $\lambda > 1$ for prolate or oblate deformations respectively. Then Heiselberg \cite{Hei04} has obtained analytical expression for the collective frequencies along the
 elongation axis or perpendicular to the elongation axis. This collective axis are called below axial or radial collective frequencies and are denoted by  $\omega_{\rm ax}$ and $\omega_{\rm rad}$ respectively.

 For prolate deformation with  $\lambda \ll 1$, the two frequencies are given by:
 \begin{eqnarray}
  \frac{\omega_{\rm ax}^p}{\omega_0}  & = & % \lambda\sqrt{3 - \frac{1}{\gamma+1}} =
  \lambda\sqrt{3 - \frac{1}{\Gamma}}, \label{eq:axp}\\
  \frac{\omega_{\rm rad}^p}{\omega_0} & = &% \sqrt{2(\gamma+1)}=
   \sqrt{2 \Gamma} , \label{eq:radp}
\end{eqnarray}
while in the oblate limit $\lambda \gg 1$, we have:
\begin{eqnarray}
  \frac{\omega_{\rm ax}^o}{\omega_0}  & = & %\lambda\sqrt{\gamma+2} =
  \lambda\sqrt{\Gamma+1},  \label{eq:axo} \\
  \frac{\omega_{\rm rad}^o}{\omega_0} & = & %\sqrt{\frac{6\gamma+4}{\gamma+2}} =
  \sqrt{\frac{6\Gamma-2}{\Gamma+1}}. \label{eq:rado}
\end{eqnarray}
Note that for $\lambda=1$ we recover results obtained for isotropic trap \cite{Coz03,Hei04}.
%In this expression, $\gamma$ is connected to the adiabatic index through the relation $\gamma \equiv \Gamma - 1$.
%\textcolor{blue}{(Denis: I am a bit puzzled by many things (i) why we do not use directly $\Gamma$ in the expression?
%(ii) In the notes, there is discussion that $\gamma$ should depend on the position which I do not understand.)}
We then see that a change in $\Gamma$ will be reflected by a change in the axial and radial collective frequencies.
\begin{figure}[htbp]
 \includegraphics[scale = 0.5]{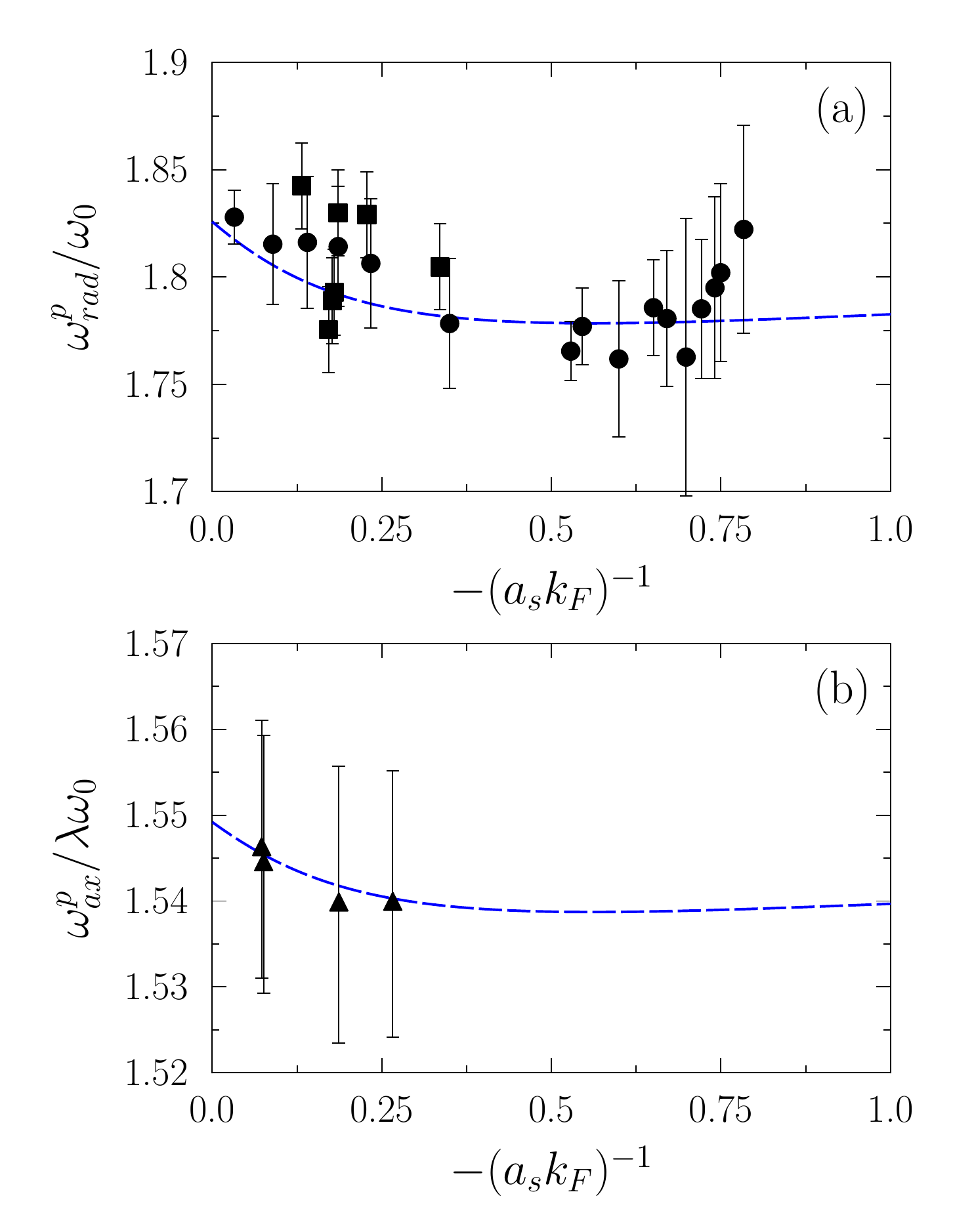}
 \caption{
 $\omega^p_{rad}/\omega_0$ (a) and $\omega^p_{ax}/(\lambda\omega_0)$ (b) as function of $-(a_sk_F)^{-1}$
 obtained with the functional \eqref{eq:func1} with $r_e = 0$ (blue dashed line).
 The symbols are experimental data: triangles from \cite{Bar04a}, circles from \cite{Kin04b}
 and squares from \cite{Kin04a}. (All data sets are taken from Fig. 3 of Ref. \cite{Man05a})}
 \label{fig:frequencies-data}
\end{figure}
The collective response of cold atoms with possible anisotropy for the trapping potential has attracted much
attention in the last decades. The experimental axial and radial frequencies are shown at
or around unitarity for prolate shapes in Fig. \ref{fig:frequencies-data}. At unitarity ($\Gamma=5/3$), we expect to have
$\omega_{\rm ax}^p / (\lambda \omega_0)=\sqrt{12/5}\simeq 1.549$ and $\omega_{\rm rad}^p / \omega_0=\sqrt{10/3}\simeq 1.826$
that seems coherent with the observations. In Fig. \ref{fig:frequencies-data}, we also display the results of Eqs. (\ref{eq:radp}) and (\ref{eq:axp}). using the adiabatic index obtained from the functional (\ref{eq:func1}) with $r_e=0$.
We see that the estimated collective frequencies are consistent with the observation in cold atoms.
We then investigate the possible effect of $r_e$ in the strict unitary regime in Fig. \ref{fig:re-dep}.
In this case, the $\Gamma$ that is used in Eqs. (\ref{eq:radp}-\ref{eq:axp}) is displayed in Fig. \ref{fig:gindexUG}.
We see a rather weak dependence of the collective frequencies with $r_e$.
\begin{figure}[htbp]
 \includegraphics[scale = 0.5]{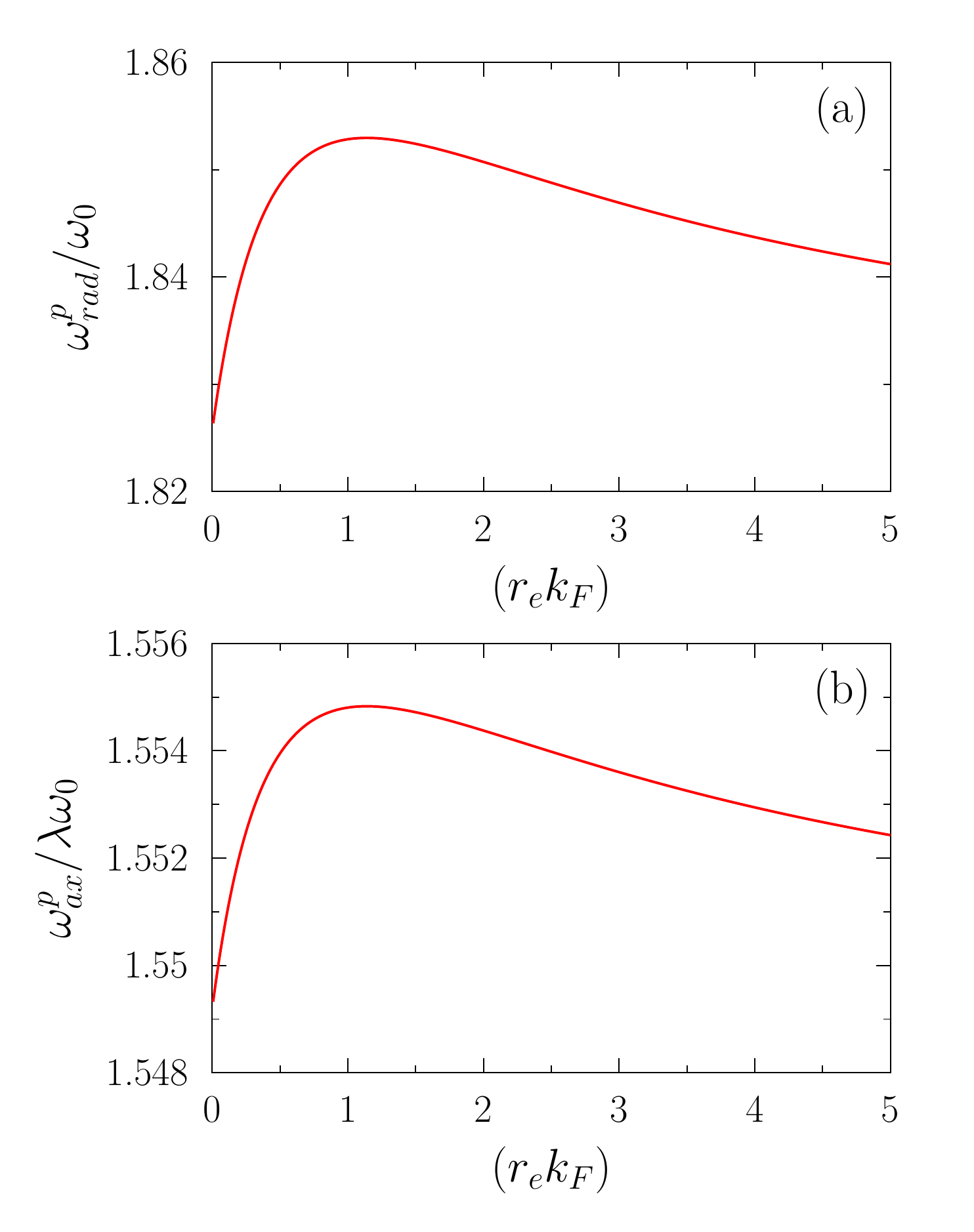}
 \caption{
 $\omega^p_{\rm rad}/\omega_0$ (a) and $\omega^p_{\rm ax}/(\lambda\omega_0)$ (b) as function of $(r_ek_F)$
 obtained with the functional (\ref{eq:func1}) at unitarity $(a_sk_F)^{-1} = 0$.}
 \label{fig:re-dep}
\end{figure}
\begin{figure}[htbp]
 \includegraphics[scale = 0.55]{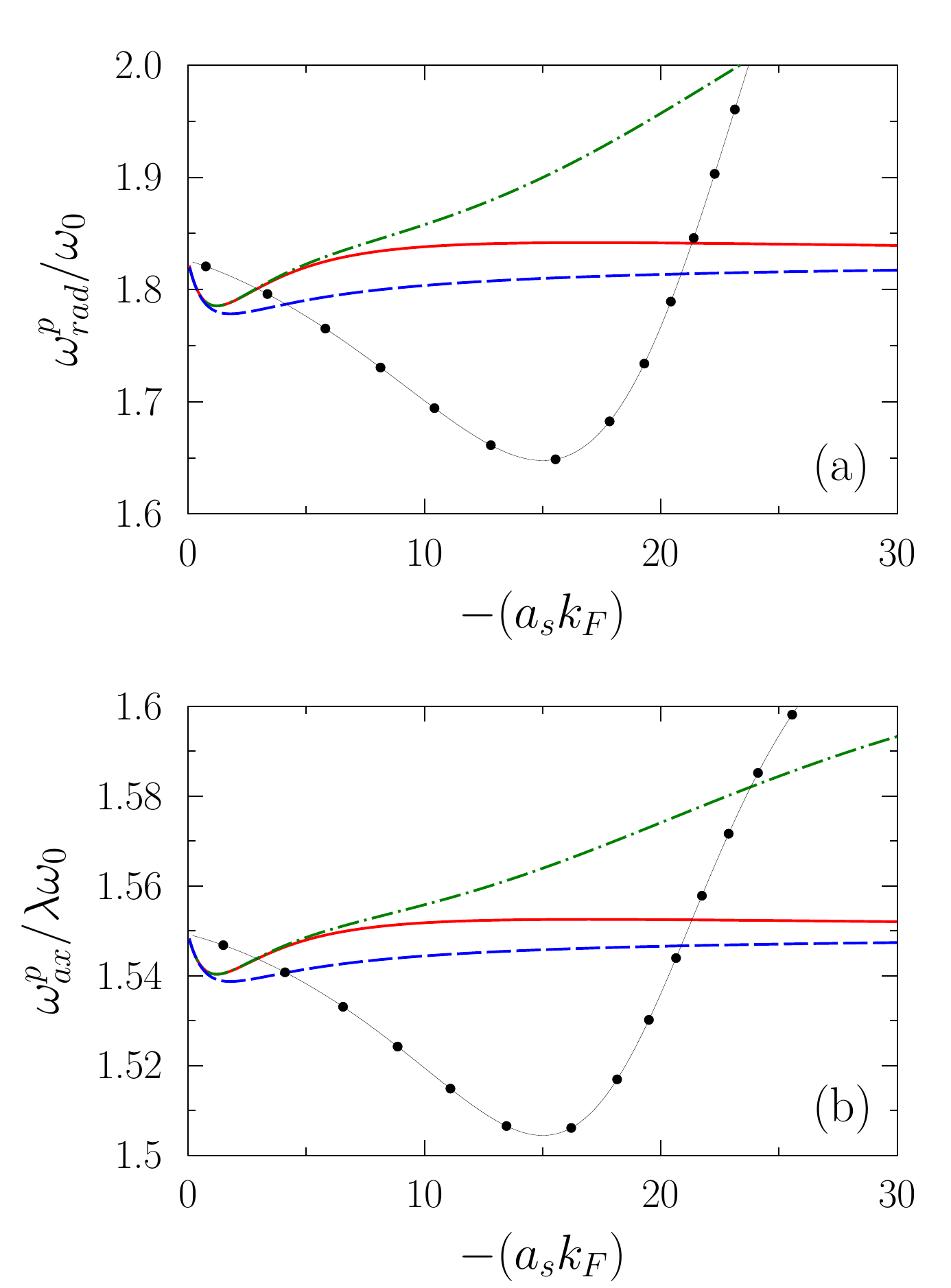}
   \caption{\label{fig:frequencies-neut-funct}
 $\omega^p_{\rm rad}/\omega_0$ (resp.  $\omega^p_{\rm ax}/\omega_0$) as a function of $-(a_sk_F)$ (a)
 (resp. (b)) obtained in neutron matter using $a_s = -18.9~\mathrm{fm}$ and
 $r_e = 2.716~\mathrm{fm}$ (red solid line) in the functional (\ref{eq:func1}) while the blue long-dashed line corresponds
 to the result obtained with $r_e=0$.
 For comparison, we also show the result of the Skyrme Sly5 parameter sets (black filled circles) and the
 result obtained by adding to the functional the leading order $p$-wave contribution (green dot-dashed line).}
\end{figure}

We finally display in Fig. \ref{fig:frequencies-neut-funct} the collective frequencies obtained for confined neutron systems in an
anisotropic trap. As far as we know, the present work is the first attempt to determine this  particular quantity
 neutronic systems.
Collective frequencies obtained with the functional are compared with the case of cold atoms and with the result of the empirical Skyrme EDF with Sly5
sets of parameter.
It is first noted that collective frequencies are strongly dependent on the used functional and therefore the dynamical  collective frequencies of trapped neutron is a stringent test of the functional used.
We finally would like to mention that the collective frequencies
are calculated here assuming that the local density approximation is valid. However, the collective frequencies might be affected by the introduction of gradients of the densities as it is usually done in more empirical functional like Skyrme ones.
 In addition, we predict  rather large differences
 between neutron matter and cold atoms that are due to effective range effects as well as higher order channels like $p$-wave when
 the density increases.

\section{Critical discussion on the role of pairing}
\label{sec:critical}

In the present article, we focused our attention on the static response of doubly degenerated Fermi liquid with anomalously large $s$-wave scattering length. We have seen that, assuming that the effective mass is approximately equal to the bare mass and neglecting possible
effect of superfluidity, our functional can describe reasonably the ground state thermodynamical quantities close or at unitarity in cold atoms and
can give interesting insight for the static response of neutron matter.
The comparison is less favorable when performing the full dynamical response. Using the same assumptions as for the static response, we also calculated the dynamical response of the system to a small oscillating external perturbation $V_{\rm ext}(\mathbf{r},t)$ with varying frequency $\omega$.
%Following Ref. \cite{Pas12}, the dynamical response can be written as (assuming $\hbar=1$):
The dynamical response function $\chi(q,\omega)$ then generalizes the static response \cite{Gar92,Pas15} that is obtained
as the specific case $\omega=0$.

One then defines the dynamical structure function
$S(q, \omega)$  through:
\begin{eqnarray}
  S(q,\omega) %\equiv \sum_{n\neq0} |\langle n|Q|0 \rangle|^2 \delta(\omega - \omega_n)
               = -\frac{1}{\pi} \Im\left[\chi(q,\omega)\right]. \label{eq:dyn}
\end{eqnarray}
While the static response function has not been directly obtained in UG, its dynamical structure function has been studied both experimentally
and theoretically  in Refs \cite{Hoi12,Ast14}.
\begin{figure}[h!]
 \includegraphics[scale=0.6]{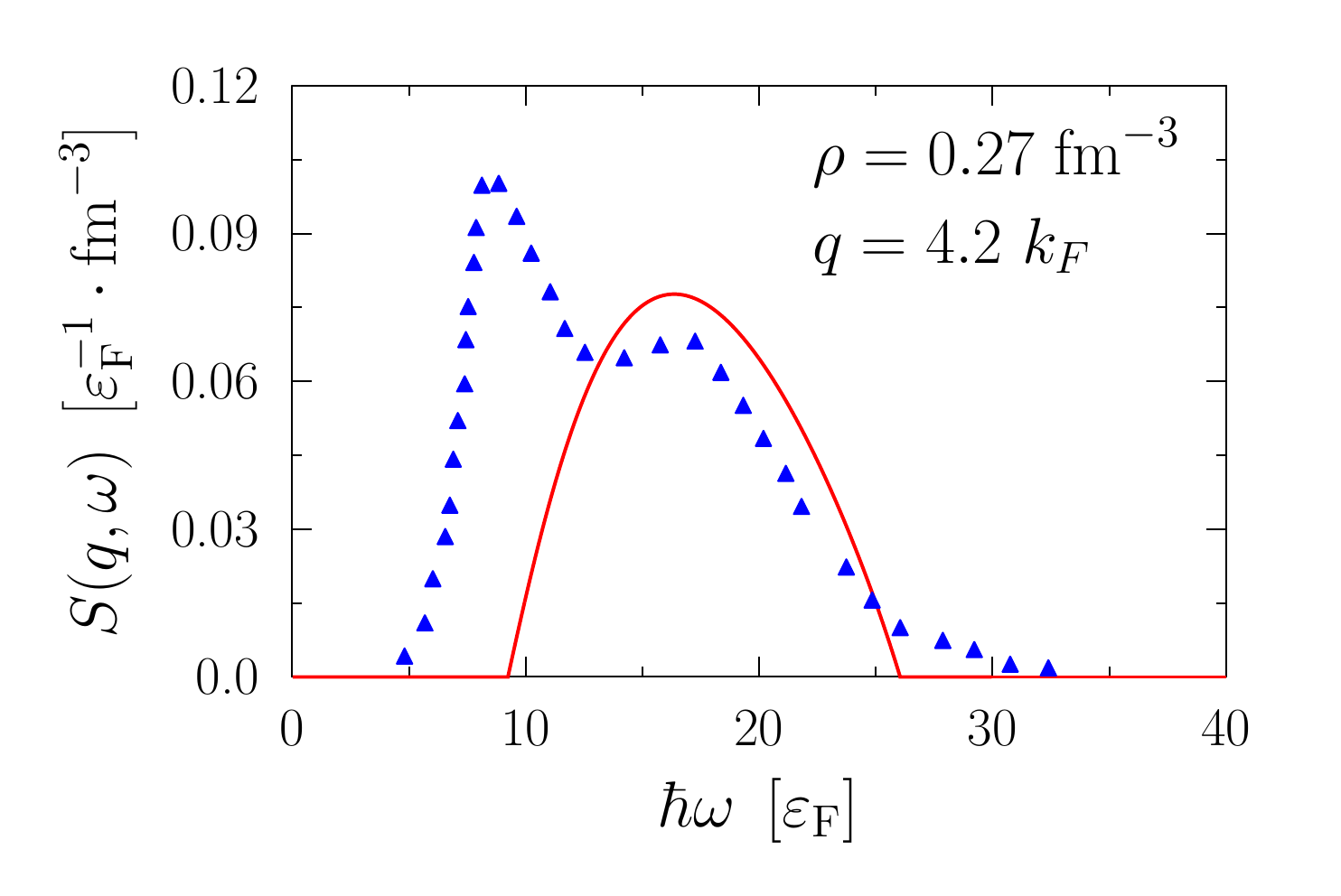}
 \caption{
 Dynamical structure function at unitarity obtained from the functional (\ref{eq:func1}) (red solid line) compared to the
 experimental results of Ref. \cite{Hoi12} (blue triangles). $\varepsilon_F = \hbar^2k_F^2/2m$ is the Fermi energy.
 }\label{fig:dyn-resp-coldatoms}
\end{figure}
The experimental structure function obtained in Ref.  \cite{Hoi12} is compared to the response obtained with the functional (\ref{eq:func1})
in Fig. \ref{fig:dyn-resp-coldatoms}.
The experimental response presents two separated peaks. We obviously see that the dynamical response obtained with our functional
is able approximately to reproduce the second peak but completely miss the collective mode at low energy. This mode is indeed due to
superfluidity leading to the so-called Bogoliubov-Anderson mode, that seems difficult to describe without explicitly
using a quasi-particle picture.  As shown in Ref. \cite{Zou16} using the RPA approach with the SLDA, accounting for superfluidity leads
back to the proper low energy collective modes that reproduces qualitatively the observation.
As shown above, many aspects can be properly reproduced in cold atoms without explicitly introducing superfluidity. However, the dynamical response
clearly points out the necessity in the near future to explicitly include the anomalous density in the description.

\begin{figure}
 \includegraphics[scale=.6]{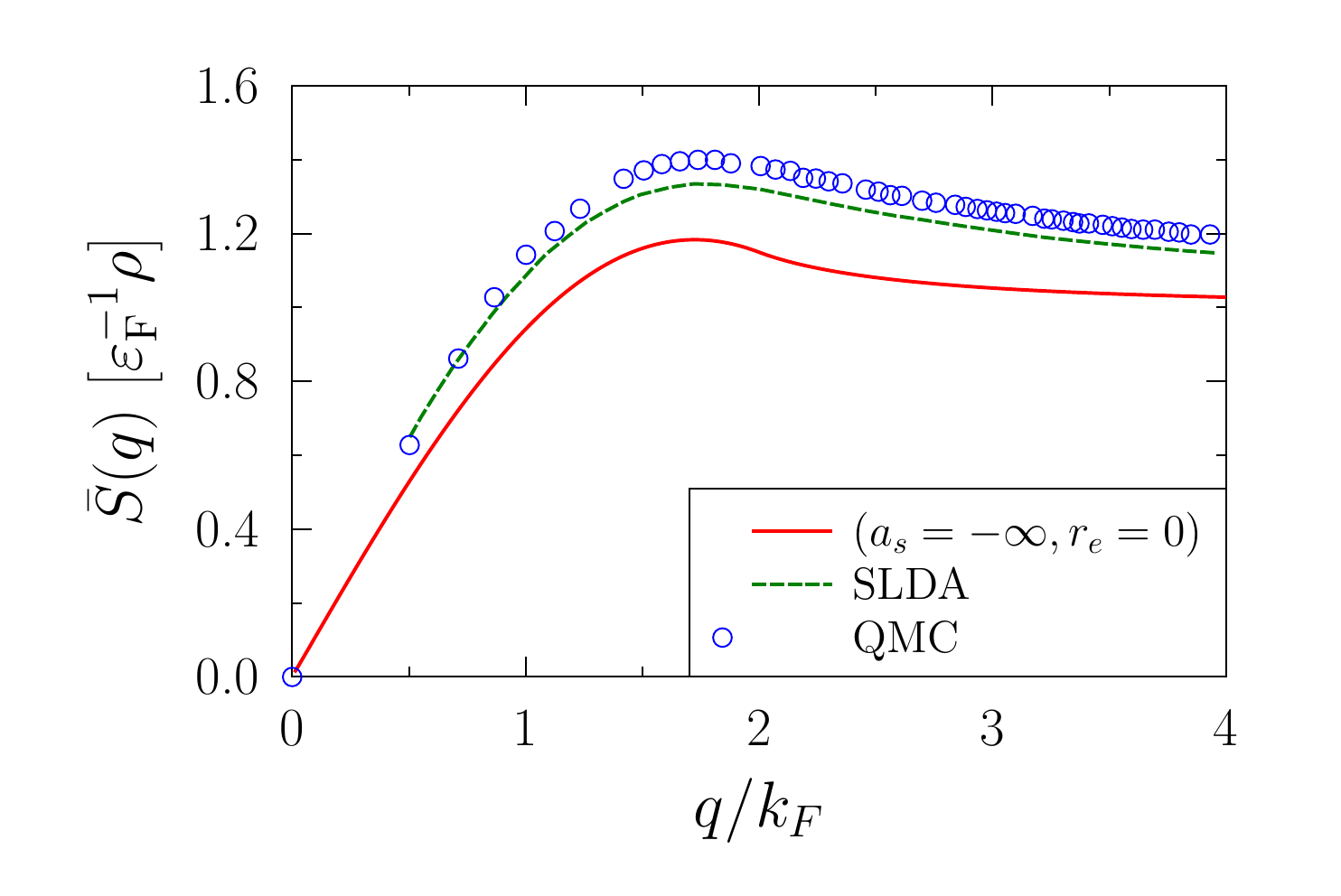}
 \caption{Static structure function obtained with the functional (with $r_e=0$) (red solid line)  as a function
 of $q/k_F$ compared to the Diffusion Monte-Carlo result of Ref.  \cite{Car14} (blue circles).
 For comparison, we also show the result obtained in Ref. \cite{Zou16} with the SLDA (green dashed line). }
 \label{fig:SUG}
\end{figure}

One can also obtain the static structure function $ \bar S(q)$, defined through
\begin{eqnarray}
\bar S(q) = \int d\omega S(q,\omega), \nonumber
\end{eqnarray}
 that has been obtained for UG in Ref. \cite{Hoi13} where it is compared
to QMC results (see also Ref. \cite{Car14}). We show in Fig. \ref{fig:SUG} a comparison of the
static structure factor obtain with the functional with the Monte-Carlo result of Ref. \cite{Car14}.
Not surprisingly, due to the missing peak at low energy, $\bar S(q)$ is underestimated compared to the exact results.
Our conclusion, is that for specific aspects like the dynamical response, it will be necessary to improve the functional
by allowing $U(1)$ symmetry breaking. The same situation will also happen for neutron matter at very low density. However,
in this case, when the density increases pairing gap exponentially decreases. In particular, at densities considered in the DFMC
results of Ref. \cite{Bur16,Bur17}, pairing is expected to not affect the static response.

\section{Conclusion}

In the present work, we make a detailed analysis of thermodynamical ground-state properties of both cold atoms and
neutron matter starting from the new density functional proposed in Refs. \cite{Lac16,Lac17}. For cold atoms with large negative $s$-wave
scattering length and with negligible effective range effects,
thermodynamical quantities like the pressure, the chemical potential, the compressibility and zero sound are very well reproduced.
We further analyze the possible influence of the effective range at and away from the unitary gas limit.
The inclusion of effective range is the first step towards the proper description of neutron matter. The difference between ground-state
thermodynamical properties  in UG and neutron matter are quantified.

The thermodynamical quantities, and more specifically the compressibility are connected to the static response of Fermi liquids to an external constraint for
which exact AFDMC exists \cite{Bur16,Bur17}. The exact static response is obtained using the new functional. It is shown to be in much better agreement with AFDMC result than the Skyrme type functional especially at low density.

We finally consider the dynamical collective response in the hydrodynamical regime. In the cold atom case, a reasonable description  of radial and axial
collective frequency is obtained assuming a polytropic equation of state. Following a similar strategy, we estimate the collective frequencies
of neutron drops in anisotropic traps. Important differences are observed between Skyrme empirical functional and the new functional discussed here.

\begin{acknowledgments}
The authors thanks J. Bonnard, A. Gezerlis, M. Grasso,  C.-Y. Yang for useful discussion at different stage of the work. D. L. also thank
A. Pastore for cross-checking and his help in obtaining the result for the response with Skyrme functional.
This project has received funding from
the European Unions Horizon 2020 research and innovation
program under grant agreement No. 654002.
\end{acknowledgments}

\end{document}